\pdfoutput=1
\documentclass[12pt,twocolumn]{IEEEtran}

\usepackage{amssymb,amsmath,graphicx,cite,epsfig,epstopdf,algorithm,algorithmic}
\usepackage[mathscr]{euscript}
\usepackage{subcaption}

\def\x{{\boldsymbol x}}
\usepackage{amsmath}

\newcommand{\y}{\mathbf{y}}

\newcommand{\R}{\mathbf{R}}

\newcommand{\A}{\mathbf{A}}
\newcommand{\bP}{\mathbf{P}}
\newcommand{\bB}{\mathbf{B}}
\newcommand{\D}{\mathbf{D}}

\newcommand{\scrE}{\boldsymbol{\mathscr{E}}}
\newcommand{\scrY}{\boldsymbol{\mathscr{Y}}}
\newcommand{\scrI}{\boldsymbol{\mathscr{I}}}
\newcommand{\scrF}{\boldsymbol{\mathscr{F}}}

\newcommand{\I}{\mathbf{I}}

\newcommand{\z}{\boldsymbol{z}}

\newcommand{\bu}{\boldsymbol{u}}

\newcommand{\bd}{\boldsymbol{d}}
\newcommand{\bp}{\boldsymbol{p}}
\newcommand{\bv}{\boldsymbol{v}}

\renewenvironment{thebibliography}[1]{%
 \begin{oldthebibliography}{#1}%
 \setlength{\parskip}{0ex}%
 \setlength{\itemsep}{0ex}%
}%
{%
\end{oldthebibliography}%
}%

\title{Estimating Sparse Signals Using Integrated Wideband Dictionaries\thanks{This work was supported in part by the Swedish Research Council and Crafoord's  foundation and in part by the ERA-chair project COEL grant (No 668995) financed by European Union’s Horizon 2020 research and innovation program, the Institutional Research Project IUT19-11 financed by the Estonian Research Council, and the project B38, financed by the Tallinn University of Technology. Parts of the material herein have been presented at the 2017 ICASSP  conference.}}

\author{Maksim Butsenko$^*$\thanks{$^*$Thomas Johann Seebeck Dept. of Electronics, Tallinn University of Technology, Ehitajate tee 5, 19086, Tallinn, Estonia, email: \texttt{maksim.butsenko@ttu.ee}.}, 
Johan Sw\"{a}rd$^\dagger$\thanks{$^\dagger$Dept. of Mathematical Statistics, Lund University, P.O. Box 118, SE-221 00 Lund, Sweden, email: \texttt{\{js,aj\}@maths.lth.se}.}, and Andreas Jakobsson$^\dagger$
}

\begin{document}
\maketitle
\begin{abstract}
In this paper, we introduce a wideband dictionary framework for estimating sparse signals. By formulating integrated dictionary elements spanning bands of the considered parameter space, one may efficiently find and discard large parts of the parameter space not active in the signal. 
After each iteration, the zero-valued parts of the dictionary may be discarded to allow a refined dictionary to be formed around the active elements, resulting in a zoomed dictionary to be used in the following iterations. Implementing this scheme allows for more accurate estimates, at a much lower computational cost, as compared to directly forming a larger dictionary spanning the whole parameter space or performing a zooming procedure using standard dictionary elements. Different from traditional dictionaries, the wideband dictionary allows for the use of dictionaries with fewer elements than the number of available samples without loss of resolution. The technique may be used on both one- and multi-dimensional signals, and may be exploited to refine several traditional sparse estimators, here illustrated with the LASSO and the SPICE estimators. Numerical examples illustrate the improved performance.
\end{abstract}
%
%


\section{Introduction}

A wide range of common applications yield signals that may be well approximated using a sparse reconstruction framework, and the area has as a result attracted notable interest in the recent literature (see, e.g., \cite{UnserT14, Elad10, CandesW08} and the references therein). Much of this work has focused on formulating convex algorithms that exploit different sparsity inducing penalties, thereby encouraging solutions that are well represented using only a few elements from some (typically known) dictionary matrix, $\D$. If the dictionary is appropriately chosen, even very limited measurements can be shown to allow for an accurate signal reconstruction \cite{CandesRT06_52, Donaho06_52}. Recently, increasing attention has been given to signals that are best represented using a continuous parameter space. 
In such cases, the discretization of the parameter space that is typically used to approximate the true parameters will not represent the noise-free signal exactly, resulting in solutions that are less sparse than desired. This problem has been examined in, e.g., 
\cite{HermanS10_4, ChiSPC11_59, StoicaB12_60}, wherein discretization recommendations and new bounds of the reconstruction guarantees were presented, taking possible grid mismatches into consideration. 
Typically, this results in the use of large and over-complete dictionaries, which, although quite efficient, often violate the assumptions required to allow for a perfect recovery guarantee.  

As an alternative, one may formulate the reconstruction problem using a continuous dictionary, such as in, e.g., \cite{TangBSR13_59, ChiC15_63, YangX16_64}. This kind of formulations typically use an
atomic norm penalty, as introduced in \cite{ChandrasekaranRPW12_12}, which allows for a way to determine the most suitable convex penalty to recover the signal, even over a continuous parameter space. 
These solutions often offer an accurate signal reconstruction, but also require the solving of large and computationally rather cumbersome optimization problems, thereby limiting the size of the considered problems. 

In this work, we examine an alternative way of approaching the problem, proposing the use of wideband dictionary elements, such that the dictionary is formed over $B$ subsets of the continuous parameter space. In the estimation procedure, the activated subsets are retained and refined, whereas non-activated sets are discarded from the further optimization. This screening procedure may be broken down into two steps. The first step is to remove the parts of the parameter space not active in the signal, whereafter, in the second step, a smaller dictionary is formed covering only the parts of the parameter space that were active in the first step. This smaller dictionary may then again be expanded with candidates close to the activated elements, thereby yielding a zoomed dictionary in these regions. The process may then be repeated to further refine the estimates as desired. Without loss of generality, the proposed principle is here illustrated on the problem of estimating the frequencies of $K$ complex-valued $M$-dimensional sinusoid corrupted by white circularly symmetric Gaussian noise. 
The one-dimensional case of this is a classical estimation problem, originally expressed using a sparse reconstruction framework in \cite{Fuchs08}, and having since attracting notable attention (see, e.g., \cite{StoicaBL11_59, StoicaB12_92,GorodnitskyR97_45,AdalbjornssonJC15_109}). Here, using the classical formulation, the resulting sinusoidal dictionary will allow for a $K$-sparse representation of frequencies on the grid, whereas the grid mismatch of any off-grid components will typically yield solutions with more than $K$ components. Extending the dictionary to use a finely spaced dictionary, as suggested in, e.g., \cite{StoicaB12_60}, will yield the desired solution, although at the cost of an increased complexity. In this work, we instead proceed to divide the spectrum into $B$ (continuous) frequency bands, each band possibly containing multiple spectral lines. 
This allows for an initial coarse estimation of the signal frequencies, without the risk of missing any off-grid components. Due to the iterative refining of the dictionary, closely spaced components are successfully separated as the dictionary is refined; as the wideband elements span the full band, no power is off-grid, avoiding the problem of a non-sparse solution due to dictionary mismatch.

Other screening methods that decrease the dictionary size have been proposed. For instance, in \cite{GhaouiVR11,tibshiraniBFHSTT12_74,XiangWR16_PP,BonnefoyERG14,FercoqGS15,LiuZWY14}, methods for finding the elements in the dictionary that corresponds to zero-valued elements in the sparse vector were  proposed. Based on the inner product between the large dictionary and the signal, a rule was formed for deeming whether or not a dictionary element was present in the signal or not. Although these methods show a substantial decrease in computational complexity, one still has to form the inner product between the likely large dictionary and the signal. To alleviate this, one may instead use the here proposed wideband dictionary elements, thereby discarding large parts of the parameter space. Since the wideband dictionary is magnitudes smaller than the full dictionary required to achieve the reconstruction, the computational complexity is significantly reduced.

The proposed principle is not limited to methods that use discretization of the parameter space; it may also be used when solving the reconstruction problem using gridless methods, such as the methods in \cite{TangBSR13_59, ChiC15_63, YangX16_64}. It has been shown that if the reconstruction problem allows for any prior knowledge about the location of the frequencies, e.g., the frequencies are located within a certain region of the spectrum, one may use this information to improve the estimates\cite{YangX16}. The proposed method may also be used to attain such prior information, and thus improving the overall estimates as a result.

To illustrate the performance of the proposed dictionary, we make use of two different sinusoidal estimators, namely the LASSO \cite{Tibshirani96_58} and the SPICE estimators \cite{StoicaBJ11_59b, StoicaZL14_33}; the first finding the estimate by solving a penalized regression problem, whereas the latter instead solves a covariance fitting problem.

The remainder of this paper is organized as follows: in the next section, the problem of estimating an $M$-dimensional sinusoidal signal is introduced, followed, in Section III, by the introduction of the proposed wideband dictionary. In Section IV, a discussion about the computational complexity reduction allowed by the proposed wideband dictionary is given, and, in Section V, the performance of the proposed wideband dictionary is illustrated by numerical examples. Finally, in Section VI, we conclude on our work.

%
\begin{figure}[t]
\includegraphics[width=3.5in, height=2.5in]{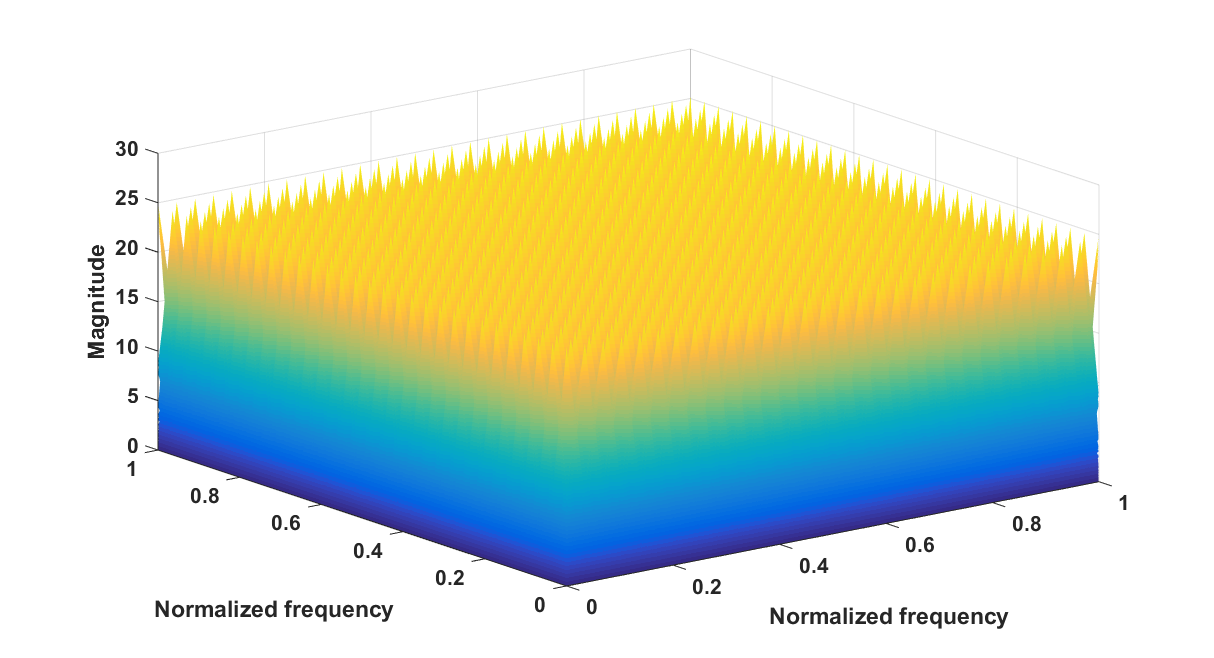}
\caption{Fine-grid dictionary for two-dimensional signal estimation with $N_1 = 30$, $N_2 = 30$, and $P = 60$ elements per dimension.} \label{fig:Dict_FineGrid}
\end{figure}

\section{Problem statement}

To illustrate the wideband dictionary framework consider the problem of estimating the $K$ frequencies $f_k^{(m)}$, for $k=1,\dots,K$ and $m=1,\dots M$, of an $M$-dimensional signal $y_{n_{1},\dots,n_{M}}$, with
%
\begin{align}\label{eq:signalModel1}
y_{n_{1},\dots,n_{M}} = \sum^K_{k=1} \beta_k e^{2i\pi f_k^{(1)} t_{n_1}^{(1)}+\dots+2i\pi f_k^{(M)} t_{n_{M}}^{(M)}} + \epsilon_{n_{1},\dots,n_{M}}
\end{align}
for $n_m=1,\ldots, N_m$, 
and where $K$ denotes the (unknown) number of sinusoids in the signal.
Furthermore, let 
$\beta_k$ and $f_k^{(m)}$ denote the complex amplitude and frequency of the $k$th frequency and $m$th dimension, respectively, $t_{n_m}^{(m)}$ the $n_m$th sample time in the $m$th dimension, and $\epsilon_{n_{1},\dots,n_{M}}$ an additive noise observed at time $t_{n_{1}},\dots,t_{n_{M}}$. The signal model in \eqref{eq:signalModel1} may be equivalently described by an $M$-dimensional ($M$-D) tensor
\begin{align}
\scrY = \sum_{k=1}^K\beta_k \tilde{\bd}^{(1)}_{(k)} \circ \tilde{\bd}^{(2)}_{(k)} \cdots \circ
\tilde{\bd}^{(M)}_{(k)}+\scrE \label{eq:signalModel2}
\end{align}
where $\circ$ denotes the outer product, and
\begin{align}
\tilde{\bd}^{(m)}_{(k)}= \left[ \begin{array}{ccc}e^{2i\pi f_k^{(m)}t^{(m)}_{1}} & \dots &
e^{2i\pi f_k^{(m)}t^{(m)}_{N_m}}\end{array}\right]^T \label{eq:params}
\end{align}
To determine the parameters of the model in \eqref{eq:signalModel1} or \eqref{eq:signalModel2}, as well as the model order, we proceed by creating a dictionary containing a set of signal candidates, each representing a sinusoid with a unique frequency. By measuring the distance between the signal candidates and the measured signal, and by promoting a sparse solution, one may find a small set of candidates that best approximates the signal. To this end, we form a dictionary on the form

\begin{align}
\D^{(m)} &=  \left[ \begin{array}{c c c} \bd_1^{(m)}  &\dots &\bd_{P_m}^{(m)} \end{array} \right] \\
\bd^{(m)}_{(p)}&= \left[ \begin{array}{ccc}e^{2i\pi f_p^{(m)}t^{(m)}_{1}} & \dots &
e^{2i\pi f_p^{(m)}t^{(m)}_{N_m}}\end{array}\right]^T
\end{align}
for $m=1,\dots, M$ and $p=1,\dots, P_{m}$, where $P_{m}\gg K$ denotes the number of candidates in dimension $m$. Here, the dictionary is assumed to be fine enough so that the unknown sinusoidal component will (reasonably well) coincide with $K$ dictionary elements\footnote{ 
As noted in \cite{ChiSPC11_59, StoicaB12_60}, the dictionary generally needs to be selected sufficiently fine to allow for a reconstruction of the signal, whereas increasing the size of the dictionary will also increase the computational complexity of the estimate. As shown in the following, the discussed method relaxes this requirement by instead defining a dictionary covering bands of potential candidates, rather than a set of individual dictionary candidates.
}.
Often, it is more convenient to work with a vectorized version of the tensor. Let $\y=\text{vec}(\scrY)$, where vec$(\cdot)$ stacks the tensor into a vector. One may then re-write \eqref{eq:signalModel2} as
\begin{align}\label{eq:vec}
\y = \left(\D^{(M)}\otimes \D^{(M-1)} \otimes \dots \otimes \D^{(1)}\right)\boldsymbol{\beta}
\end{align}
where $\otimes$ denotes the Kronecker product, suggesting that one may find both the unknown parameters and the model order by forming the LASSO problem (see, e.g.,  \cite{Fuchs08,Tibshirani96_58})
\begin{align}\label{eq:LASSO}
\underset{\boldsymbol{\beta}}{\text{min}}\ ||\y-\D\boldsymbol{\beta}||_2^2+\lambda ||\boldsymbol{\beta} ||_1
\end{align}
where $\D=\left(\D^{(M)}\otimes \D^{(M-1)} \otimes \dots \otimes \D^{(1)}\right)$ and $\|\mathord{\cdot}\|_q$ denotes the $q$-norm. A visual representation of such dictionary is shown in Figure~\ref{fig:Dict_FineGrid} for the $2$-D case. The penalty on the 1-norm of $\boldsymbol\beta$ will ensure that the found solution will be sparse, with $\lambda$ denoting a user parameter governing the desired sparsity level of the solution. The frequencies, as well as their order, are then found as the non-zero elements in $\boldsymbol \beta$.

%
\begin{figure}[t]
\includegraphics[width=3.5in, height=2.5in]{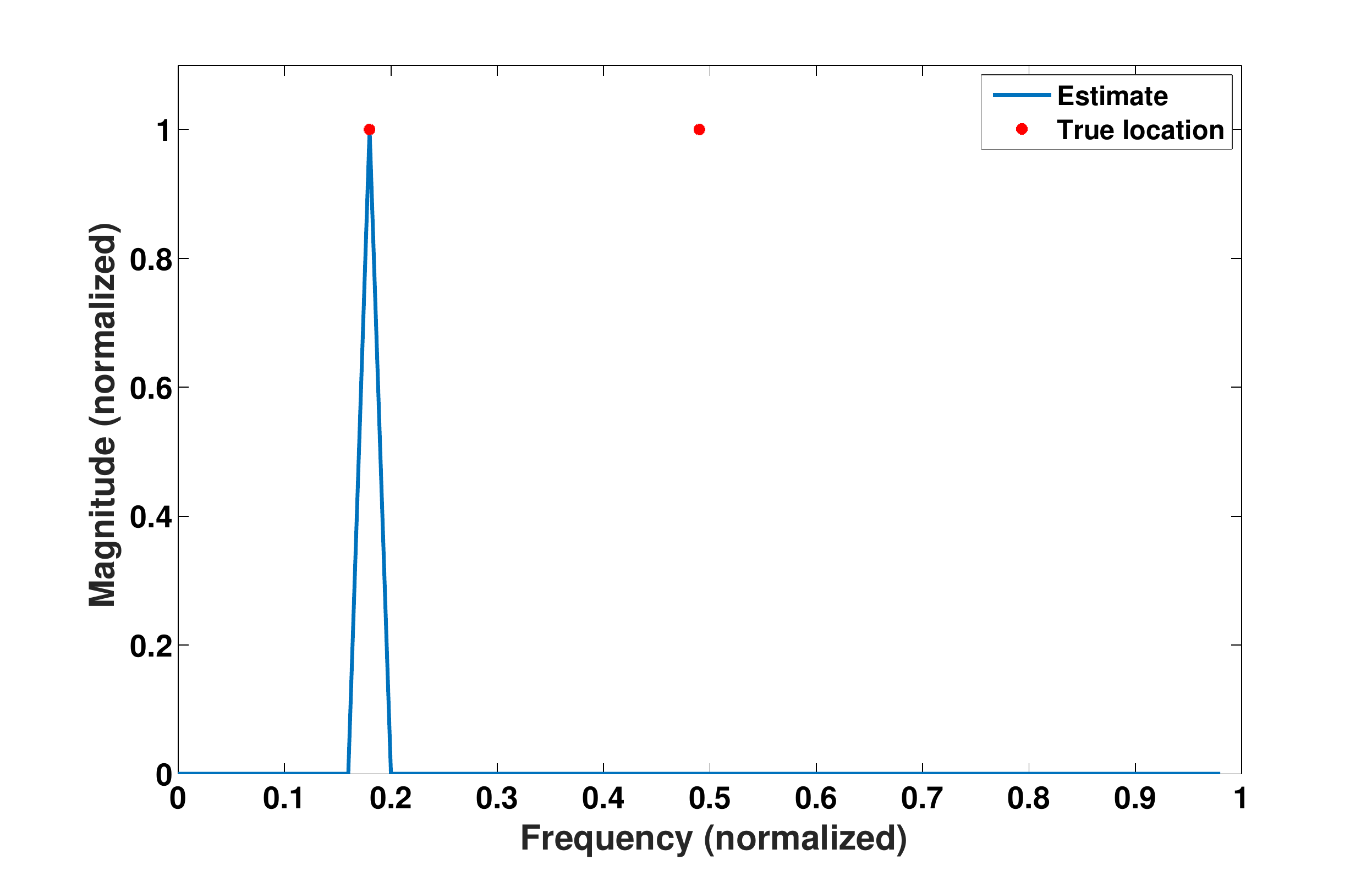}
\caption{The inner-product of a dictionary containing \mbox{$P=50$} (narrowband) candidate frequency elements and the noise-free signal, with $N=100$.} \label{fig:sinusoidBetween}
\end{figure}
%

\begin{figure}[t]
\includegraphics[width=3.5in, height=2.5in]{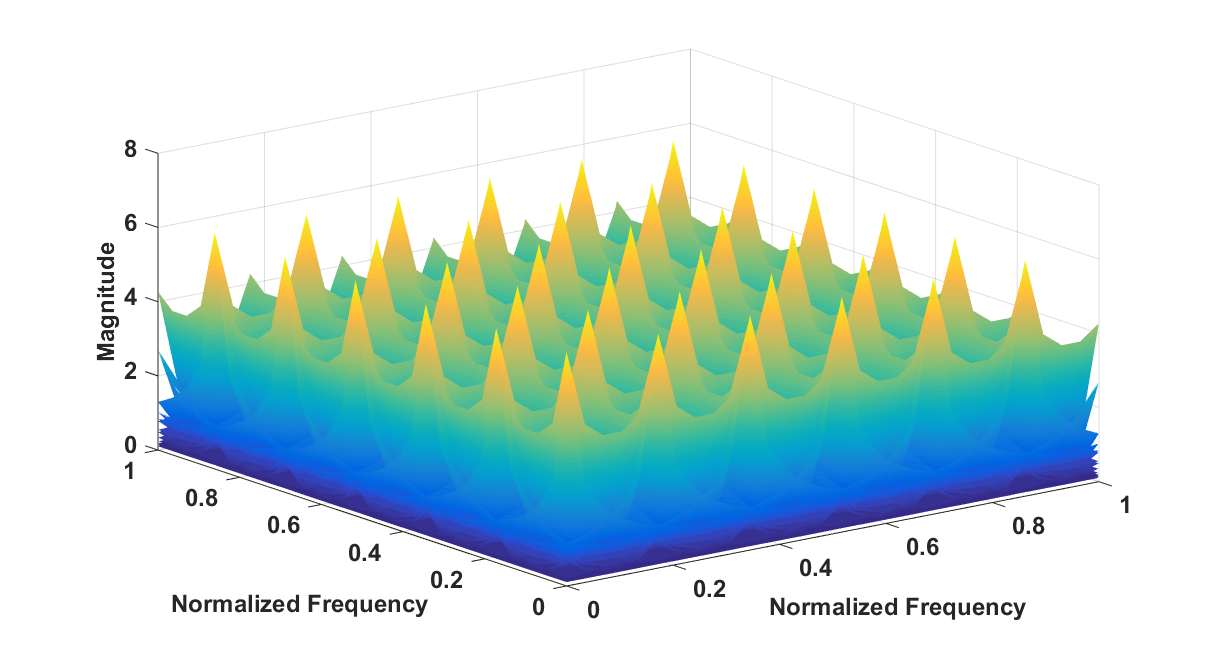}
\caption{Wideband dictionary with integrated sinusoids elements formed with $N_1 = 30$, $N_2 = 30$, and $B = 6$ bands per dimension.} \label{fig:Dict_IntSin}
\end{figure}

As shown in \cite{StoicaB12_60}, the number of dictionary elements, $P$, typically has to be large to allow for an accurate determination of the correct parameters. This means that for multi-dimensional signals, the dictionary quickly becomes inhibitory large. Thus, it is often not feasible in practice to directly compute the solution of \eqref{eq:LASSO} using a dictionary constructed from such finely space candidates.
As an alternative, one may use a zooming procedure, where one first employs an initial coarse dictionary, $\D_1$, to determine the parameter regions of interest, and then employ a fine dictionary, $\D_2$, centered around the initially found candidates (see, e.g., \cite{SahnounDB13, SwardAJ16_128} for similar approaches). This allows for a computationally efficient solution of the optimization problem in  \eqref{eq:LASSO}, but suffers from the problem of possibly missing off-grid components far from the  initial coarse frequency grid.
This is illustrated in Figure~\ref{fig:sinusoidBetween} for a $1$-D signal, where the inner-product between the dictionary and the signal is depicted together with the location of the true peaks. In this noise-free example, we used $N=100$ samples and $P=50$ dictionary elements, with one of the frequencies being situated in between two adjacent grid points in the dictionary. As seen in the figure, the coarse initial estimate fails to detect the presence of the second signal component, which is thereby discarded as a possibility in the following refined estimate. 
Increasing the number of candidate frequencies will result in the side-lobes of the dictionary elements  decreasing the gap between the frequency grid points, making the inner-product between the dictionary and the signal larger for components that lie in between two candidate frequencies. However, doing so will increase the computational complexity correspondingly, begging the question if one may retain a low number of candidate frequencies, while still reducing the likelihood of missing any off-grid components. This is the problem we shall examine in the following.

\begin{figure}[t]
\includegraphics[width=3.5in, height=2.5in]{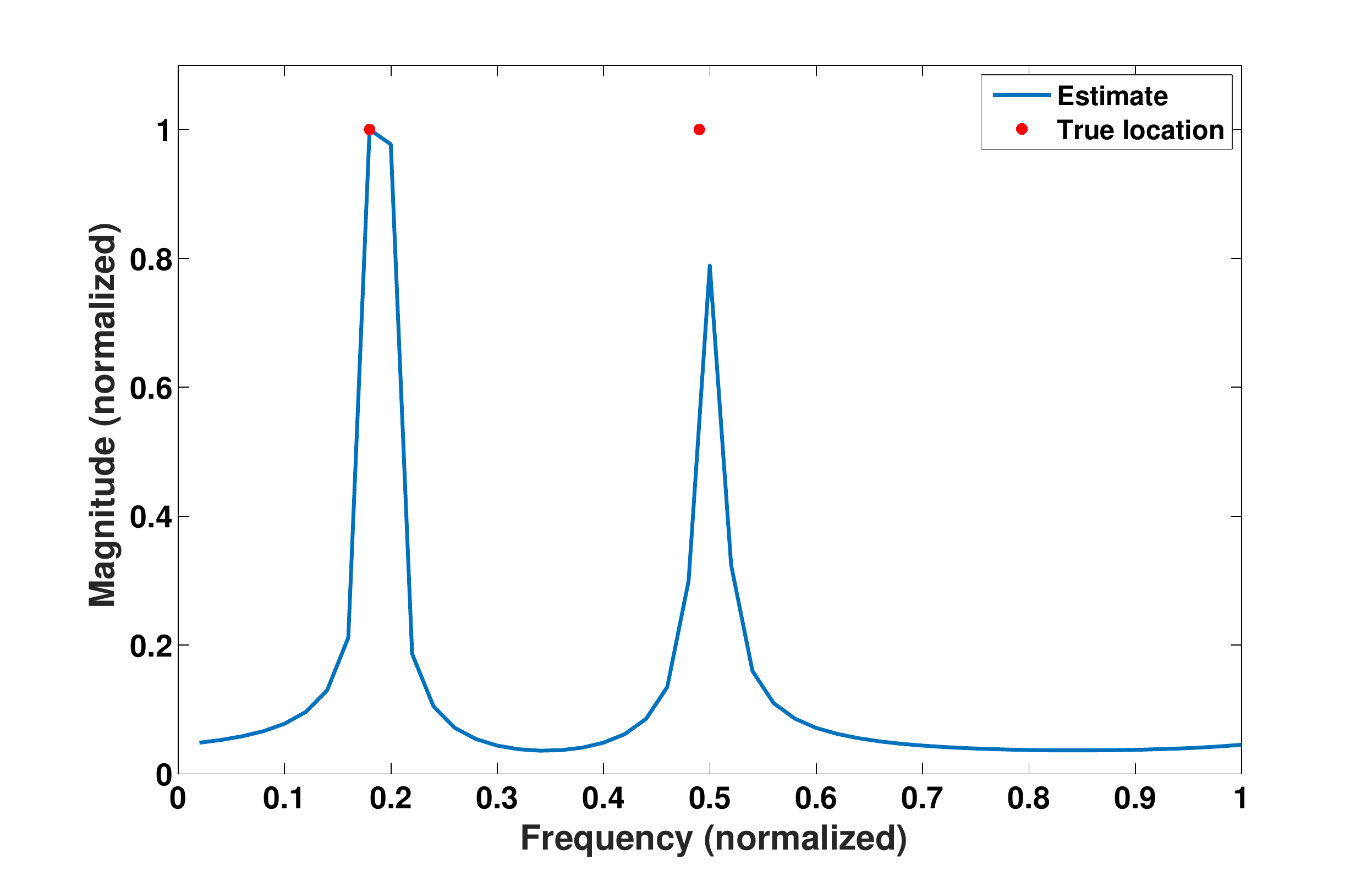}
\caption{The inner-product of a dictionary containing \mbox{$B=50$} (wideband) candidate frequency elements and the noise-free signal, with $N=100$.} \label{fig:sinusoidBetweenBL}
\end{figure}

\section{Integrated Wideband dictionaries}
We note that the above problem results from the dictionary being formed over a set of single-component candidates, thereby increasing the risk of neglecting the off-grid components. In order to avoid this, we here propose a wideband dictionary framework, such that each of the dictionary elements is instead formed over a range of such single-component candidates. This is done by letting the dictionary elements be formed over an integrated range of the parameter(s) of interest, in this case being the frequencies of the candidate sinusoids. For a multi-dimensional sinusoidal dictionary, the resulting $B$ integrated wideband elements should thus be formed as
\begin{align}\label{eq:prop}
&a_{b^{(1)},\dots,b^{(M)}}(t^{(1)},\dots,t^{(M)}) =\nonumber\\ &\int_{f_{b^{(1)}}}^{f_{b^{(1)}+1}}\cdots \int_{f_{b^{(M)}}}^{f_{b^{(M)}+1}}\nonumber\\ &e^{2i\pi\left( f^{(1)} t^{(1)}+\dots+f^{(M)} t^{(M)}\right)}df^{(1)}\dots df^{(M)}
\end{align}
for $t^{(m)}=1,\dots,N_m$ for all $m=1,\dots,M$, where $f_{{b}^{(m)}}$ and $f_{{b}^{(m)} +1}$ are the two frequencies bounding the frequency band, for $b=1,\dots,B$, for the $m$th dimension. The resulting elements are then gathered into the dictionary, $\bB$, where each column contains a specific wideband of the $M$-D parameter space for all time samples, where each element is formed as the solution from \eqref{eq:prop}, such that, in this case, 
\begin{align}\label{eq:newDictionaryElement}
a_{b^{(1)},\dots,b^{(M)}}(t^{(1)},\dots,t^{(M)}) =\nonumber\\ \prod_{m=1}^{M}\frac{e^{2i\pi f_{b^{(m)} +1} t^{(m)}}-e^{2i\pi f_{b^{(m)}} t^{(m)}} }{2i\pi t^{(m)}}
\end{align}
%
\begin{figure}[t]
\includegraphics[width=3.5in, height=2.5in]{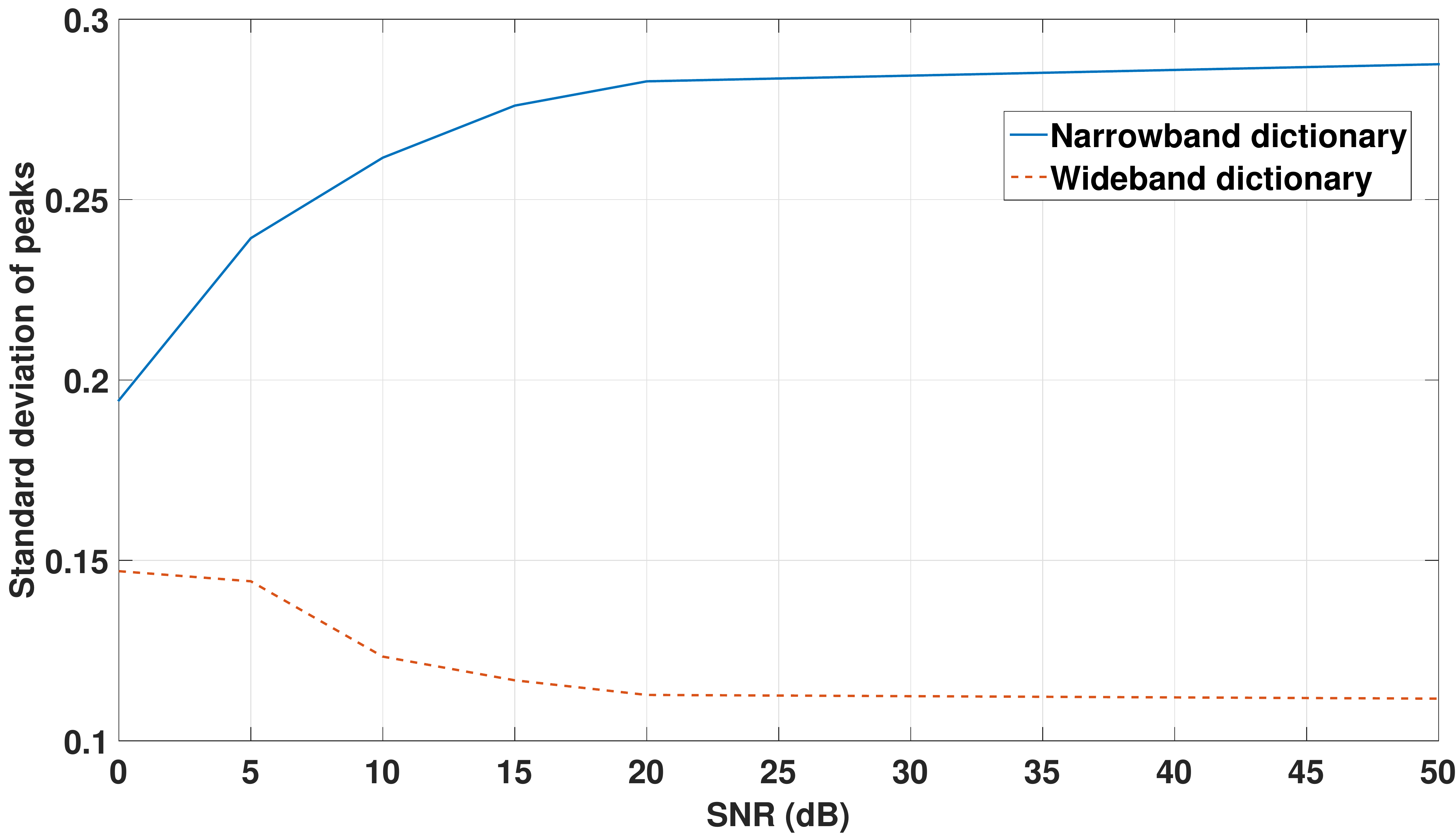}
\caption{The standard deviation of the peaks as a function of SNR.} \label{fig:peakVariance}
\end{figure}
Note that \eqref{eq:newDictionaryElement} corresponds to the $M$-D inverse Fourier transform of $1$, i.e., it is the $M$-D inverse Fourier transform of an $M$-D section in the frequency domain with unit amplitude. For the $1$-D case, this simplifies to 

\begin{align}
\begin{cases}
1,\ \text{  for }\ f_a\leq f\leq f_b \\ 
0
\end{cases}
\xrightarrow{\mathscr{F}^{-1}}
\frac{e^{2i\pi f_{b}t}-e^{2i\pi f_{a} t }}{2i\pi t}
\end{align}
Algorithm~\ref{Alg1} summarizes the usage of the wideband dictionary in a sparse reconstruction framework. In Figure~\ref{fig:Dict_IntSin}, we show a visual representation of the resulting wideband dictionary for $M = 2$ dimensions. 
The inner-product between the proposed dictionary, $\bB$, and the earlier $1$-D signal is shown in Figure~\ref{fig:sinusoidBetweenBL}, using the same number of dictionary elements as in Figure~\ref{fig:Dict_FineGrid}, clearly indicating that the proposed dictionary is able to locate the off-grid frequency. 
This is due the wideband nature of the proposed dictionary, which thus has less power concentrated at the grid points, but covers a wider range of frequencies, not reducing to zero, or close to zero, anywhere within the band (as is the case for the narrowband dictionary elements).  
%
%
%
As a result, using the wideband dictionary elements, it is possible to use a smaller dictionary, thereby reducing the computational complexity, without increasing the risk of missing components in the signal. 
To further show this, $1000$ Monte-Carlo simulations were conducted for each considered signal to noise ratio (SNR), here defined as 
\begin{align}
\text{SNR = 10log}_{10}\Big(\frac{P_y}{\sigma^2}\Big)
\end{align}
where $P_y$ is the power of signal, and $\sigma^2$ the variance of the noise.
In each simulation, we considered a signal containing two sinusoids, where the frequencies were randomly selected on $(0,1]$ with a spacing of at least $2/N$, with $N=100$ denoting the signal length. 
The sinusoids had the magnitudes $4$ and $5$, with a randomly selected phase between $(0,2\pi]$. Two dictionaries were given, one containing ordinary sinusoids and one containing the proposed wideband components, both containing $P=B=50$ elements. For each dictionary, the inner-products with the signal where computed, where the amplitudes were normalized so that the largest estimated peak had unit magnitude.  
Figure \ref{fig:peakVariance} shows the variance of the smallest peak for different SNR-levels. As is clear from the figure, the variance of the peaks are much lower for the banded case. The reason why the sinusoidal dictionary results in a larger variance is due to the fact that the main lobe is much thinner in this case than in the banded counterpart. This means that when the sinusoids happen to have frequencies that do not overlap with the main lobe of the dictionary, the power in the inner-product will be small. This will not only make such components harder to detect, but will also make it more difficult to determine a suitable regularizing hyperparameter, $\lambda$.
\begin{algorithm}[t]
\caption{Sparse reconstruction with LASSO using the wideband dictionary for the $1$-D case}
\label{Alg1}
\begin{algorithmic}[1]

\STATE choose the number of zooming steps, $I_{zoom}$
\STATE choose the number of bands, $B_1$
\STATE set the frequency bin  $\Delta_1 = \frac{1}{B_1}$
\STATE $\scrF_1 = \{f_k: f_k = k\Delta_1, \text{for } k = 1, \dots, B_1\}$
\STATE form the dictionary $\bB_1$ according to (\ref{eq:newDictionaryElement})
\STATE solve $\underset{\boldsymbol{\beta_1}}{\text{min}}\ ||\y-\bB_1\boldsymbol{\beta_1}||_2^2+\lambda ||\boldsymbol{\beta_1} ||_1$
\STATE $\scrI_1 = \{i: \beta_1(i) > 0, \text{for } i = 1, \dots, B_1\}$
\STATE $\scrF_{active} = \{f_k \in \scrF_1: k \in \scrI_1\}$
\FOR[zooming procedure] {$z=2$ to $I_{zoom}$}
	\STATE choose the number of bands, $B_z$
	\STATE select the frequency bin  $\Delta_z = \frac{\Delta_{z-1}}{B_z}$
	\STATE $\scrF_z = \{\mathbf{f}_k: \mathbf{f}_k = [f_k + \Delta_z, f_k + 2\Delta_z, \dots, f_k + B_z\Delta_z]^T, f_k \in \scrF_{active}\}$
	\STATE form the dictionary $\bB_z$ according to (\ref{eq:newDictionaryElement})
	\STATE solve $\underset{\boldsymbol{\beta_z}}{\text{min}}\ ||\y-\bB_z\boldsymbol{\beta_z}||_2^2+\lambda ||\boldsymbol{\beta_z} ||_1$
	\STATE $\scrI_z = \{i: \beta_z(i) > 0, \text{for } i = 1, \dots, \prod_{1}^{z}B_z\}$
	\STATE $\scrF_{active} = \{f_k \in \scrF_z: k \in \scrI_z\}$
\ENDFOR

\end{algorithmic}
\end{algorithm}

When $P$ decreases below $N$, the gaps between the frequency candidates in the single-component dictionary become so large that if one of the sinusoids in the signal has its frequency values between two adjacent grid points, the likelihood that this sinusoid lie in the null-space of the dictionary increases. This problem is avoided with the wideband dictionary as it is more likely to eliminate any gaps.
\begin{figure*}
\begin{subfigure}[l]{0.48\textwidth}
\includegraphics[width=\textwidth]{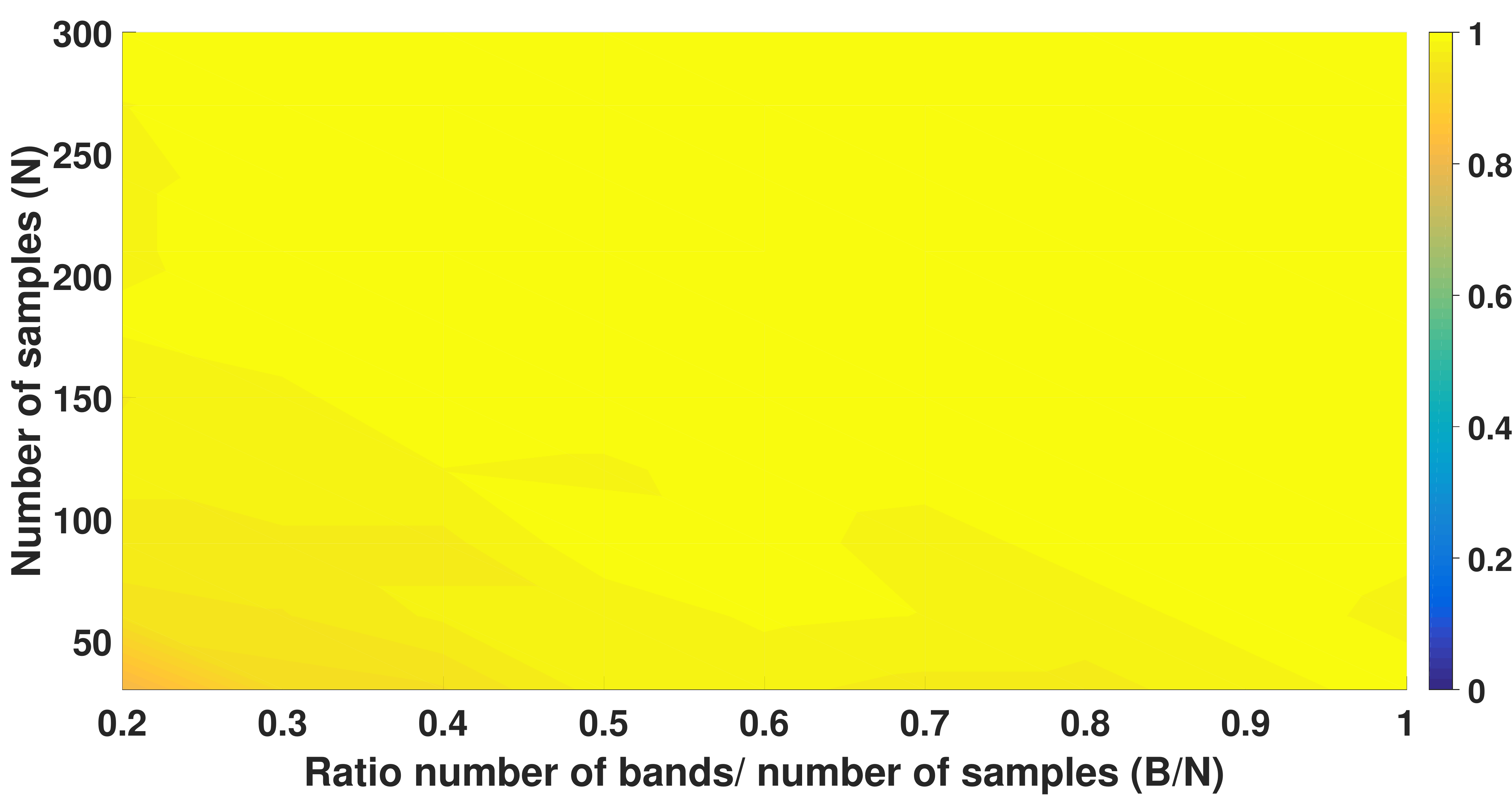}
\end{subfigure}
\begin{subfigure}[c]{0.48\textwidth}
\includegraphics[width=\textwidth]{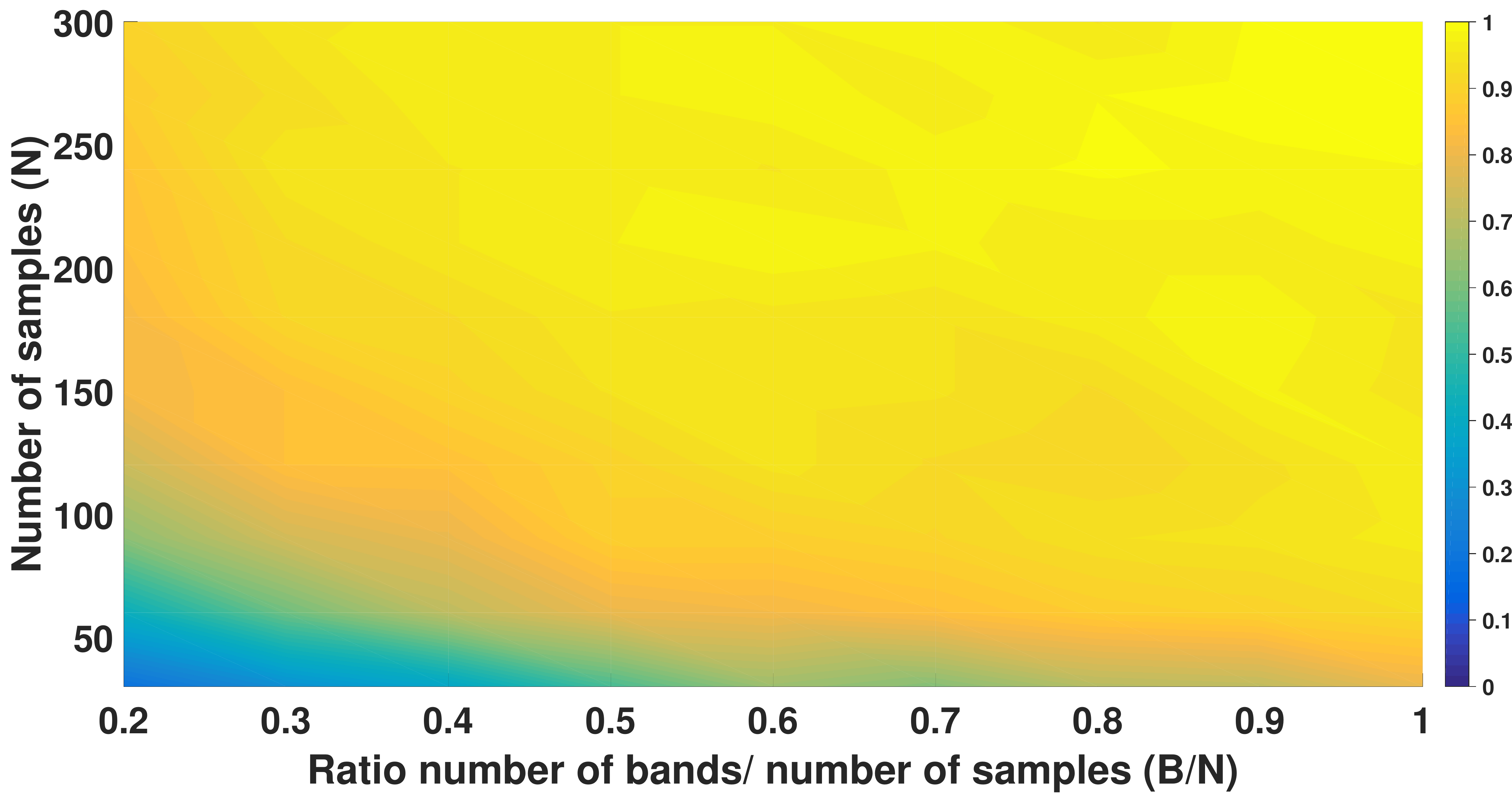}
\end{subfigure}
\centering
\begin{subfigure}[c]{0.48\textwidth}
\includegraphics[width=\textwidth]{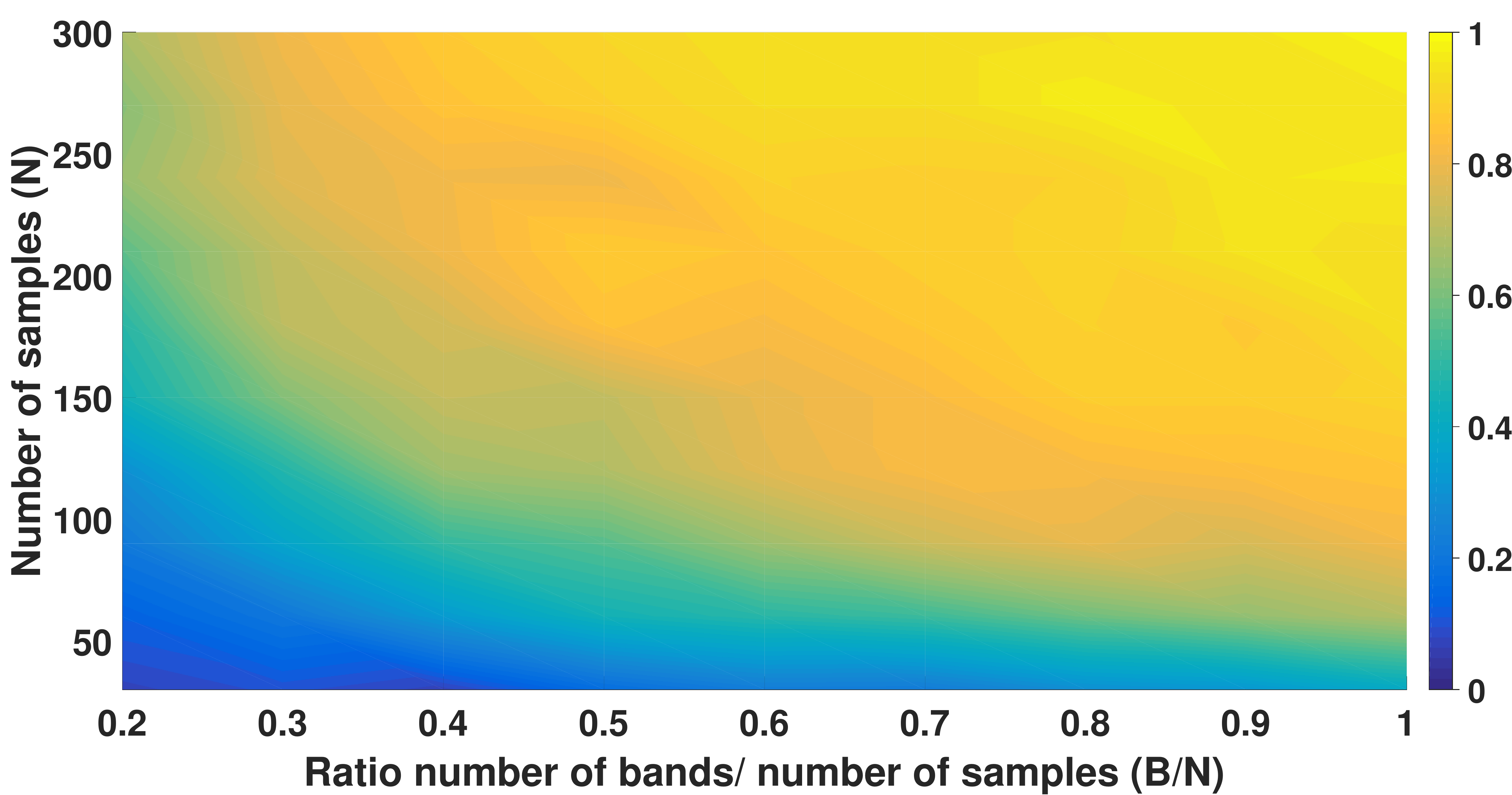}
\end{subfigure}
\caption{The success rate of finding the true support as a function of the number of samples (y-axis) and the ratio between the number of bands in the dictionary and the number of samples (x-axis), for different values of $K$. Top left corner, $K=3$, top right corner, $K=7$, and bottom, $K=11$.}\label{fig:breakDown}
    \end{figure*}

This property is depicted in Figure \ref{fig:breakDown}, where the success rate of finding the true support is displayed as a function of the number of samples, $N$, and the number of bands in the dictionary, $B$, for different number of sinusoids in the signal, $K$. The estimation was done for a noise-free signal by solving \eqref{eq:LASSO}, using wideband dictionaries and letting
\begin{align}
\lambda=0.3\underset{i=1,\dots,B}{\max}|\bd_i^T\y|
\end{align}
where $\bd_i$ denotes the $i$th column of $\D$ and the coefficient $0.3$ is selected given the observations in Figure~\ref{fig:guesses}. For a more complete discussion on how one should select $\lambda$, we refer the reader to the original presentation of the LASSO \cite{Tibshirani96_58}. In the top left figure, the signal contains three sinusoids, and it is clearly the case that the banded dictionary is able to retrieve the true support for all setting of $N$ and $B/N$, except for the case when $N=30$ and $B/N<7$. In the top right and bottom figures, where $K=7$ and $K=11$, respectively, it is shown that when the number of sinusoids in the signal increases, a larger number of samples is needed to allow for a successful reconstruction, which is reasonable, as one needs more information to be able to correctly estimate more parameters. 
However, the banded dictionary is able to retrieve the true support as long as the number of samples is big enough and the ratio $B/N$ is not too small. It is further clear from the figures, that the banded dictionary actually retrieves the true support even though $B<N$. 

The proposed approach is not the only way to form a wideband dictionary. For example, one could populate the dictionary using discrete prolate spheroid sequences (DPSS) \cite{Slepian78_57}. For an integer $Q$ and with real-valued $0 < W < \frac{1}{2}$, the DPSS are a set of $Q$ discrete-time sequences for which the amplitude spectrum is band-limited. The most interesting property of the DPSS for our discussion is the fact that the energy spectrum of the dictionary elements are highly concentrated in the range $[-W, W]$, suggesting that the DPSS could be a suitable basis for the candidates in a wideband dictionary, where the candidates are formed such that each covers a $1/B$-th part of the spectrum. In the numerical section below, we examine how the use of DPSS candidates compare to the integrated wideband candidates in \eqref{eq:newDictionaryElement}.

It is worth stressing that the wideband dictionary framework introduced here is not limited to the LASSO-style minimizations such as the one examined in \eqref{eq:LASSO}. There are many other popular methods that could be implemented using this approach. As an example of how the wideband dictionary can be applied for other typical sparse estimation algorithms, consider the SPICE algorithm \cite{StoicaBL11_59, StoicaZL14_33}, formed as the solution to
\begin{align}\label{eq:SPICE}
\underset{\tilde\bp\geq 0}{\text{minimize}}\ \y^*\R^{-1}\y+||\bp||_1+||\boldsymbol{\sigma} ||_1
\end{align}
where
\begin{align}\label{eq:R}
\R(\tilde\bp) &= \A\bP\A^* \\
\A &=\left[\begin{array}{c c c} \bB & \mathbf{I}\end{array}
\right] \\ \label{eq:pDef}
\bp &=  \left[\begin{array}{c c c c c c} p_1 & \dots & p_M \end{array} \right]^T\\\label{eq:sigmaDef}
\boldsymbol{\sigma} &= \left[\begin{array}{c c c} \sigma_1& \dots & \sigma_N
\end{array} \right]^T\\
\tilde \bp &= \left[\begin{array}{c c c} \bp^T & \boldsymbol{\sigma}^T
\end{array} \right]^T\\
\bP &= \text{diag}\left(\tilde \bp\right)
\end{align} 
Alternatively, one may consider the more general $\{r,q\}$-SPICE formulation\footnote{In this formulation, we assume that the columns of the dictionaries are normalized to have unit norm.} \cite{SwardAJ_icassp17,SwardAJ17_accepted}
\begin{align}\label{eq:rq_SPICE}
\underset{\tilde\bp\geq 0}{\text{minimize}}\ \y^*\R^{-1}\y+||\bp||_r+||\boldsymbol{\sigma} ||_q
\end{align}
Using the wideband dictionary over $\bB$ in \eqref{eq:SPICE} or \eqref{eq:rq_SPICE} will allow for much smaller dictionaries as opposed to using ordinary sinusoidal dictionaries.  Many other sparse reconstruction techniques may be extended similarly. Generally, the wideband dictionary may be used either as an energy detector which finds the parts of the spectrum that have most energy, or in a zooming procedure similar to the one described above. 

%

\begin{figure}[t]
\includegraphics[width=3.5in, height=2.5in]{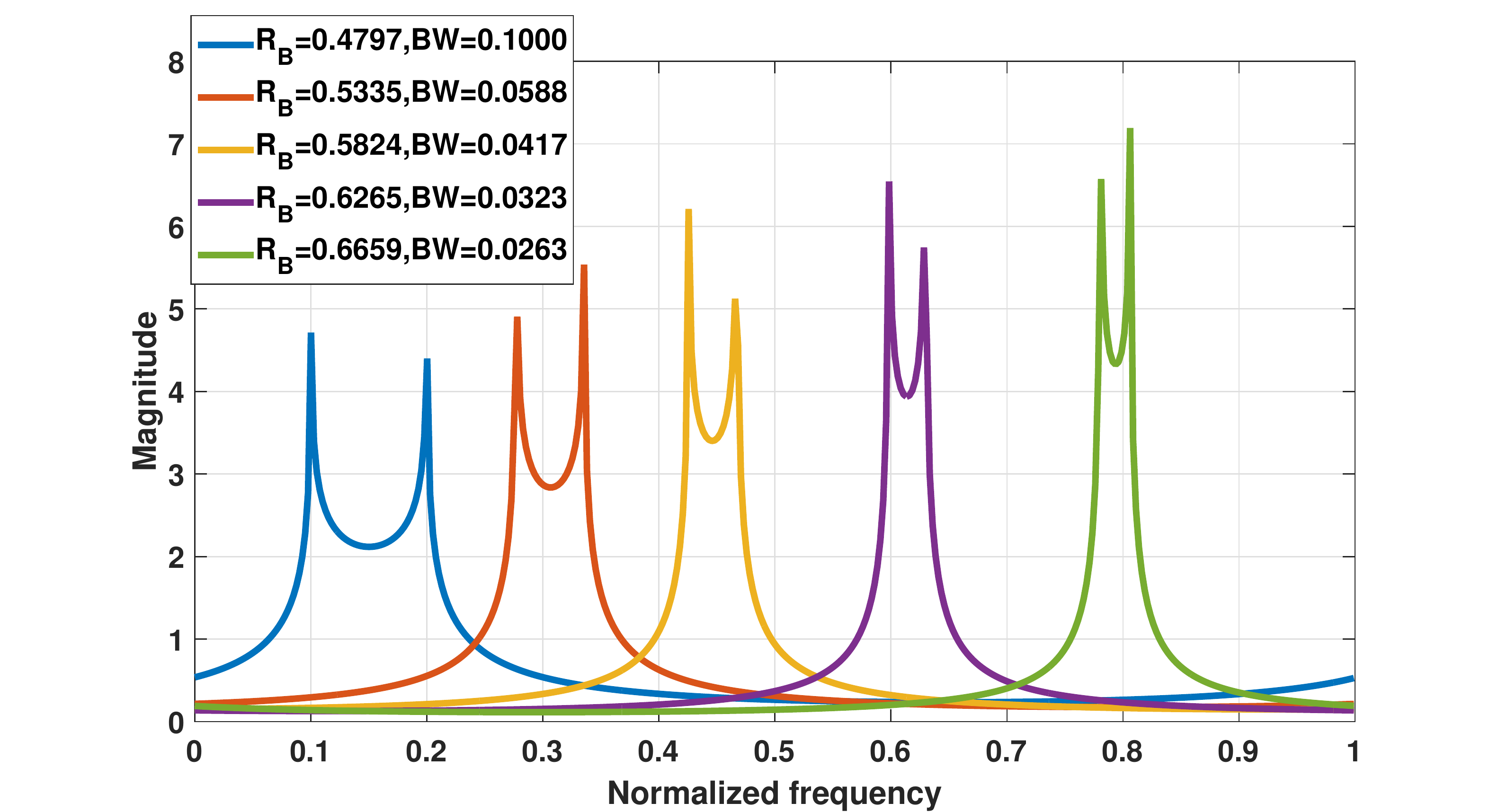}
\caption{A single wideband element for varying bandwidths ($R_B$ - ratio, $BW$ - bandwidth).} \label{fig:onebandelement}
\end{figure}

\section{Parameter Selection}

From our discussion on the integrated wideband dictionary and its use for sparse signal estimation, one may note that there are two parameters which should be chosen by the user, namely the number of used bands and the number of zooming steps. The choice of the number of the bands, $B$, will depend on the required resolution, whereas the number of zooming steps will decrease the computation complexity (for a fixed resolution) as with each zooming step inactive parts of the spectra are discarded from  future computations. 
Therefore, the choice of the total number of bands one should use is dependent on the required resolution. Furthermore, for each zooming step, the distribution of these bands should be made such that the subsequent selection will guarantee a high likelihood of including the true support. 
This idea is illustrated in Figure~\ref{fig:breakDown}, where the success rate of finding the true support is shown to depend on the number of bands, the number of samples, and the number of components in the data.

As may be expected, the use of the wideband dictionary does not remove such user choices; in fact, the here proposed framework does not remove any of the usual user choices or limitations of a sparse reconstruction technique, be it the LASSO, SPICE, or any other dictionary based technique, and the same restrictions will apply that do so for the particular method if used with a narrowband dictionary. Rather, the wideband dictionary allows for an efficient refinement procedure speeding up the calculations required in forming the estimate.

\begin{figure}[t]
\includegraphics[width=3.5in, height=2.5in]{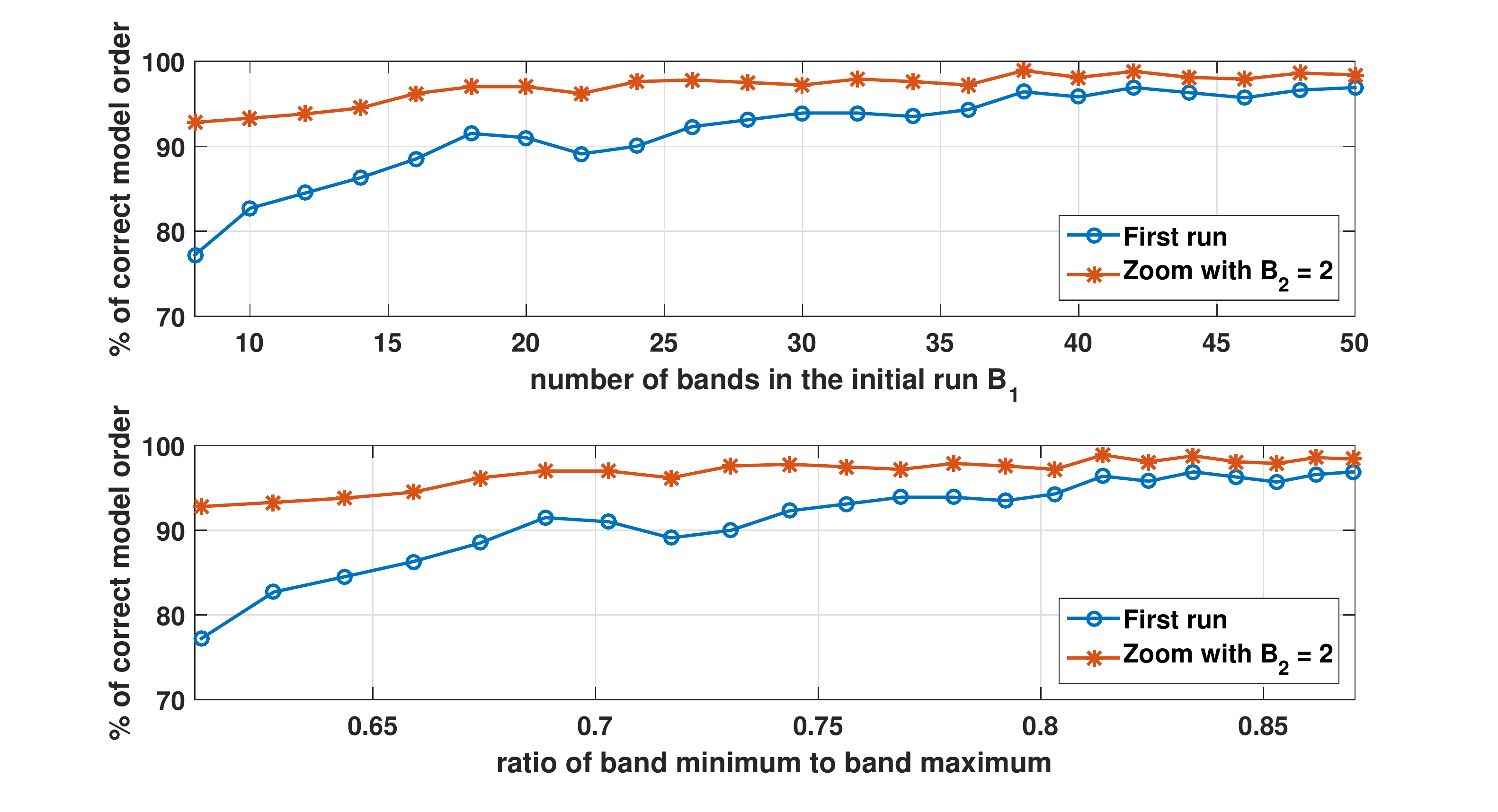}
\caption{Percentage of finding correct model order as a function of number of bands(top) and ratio of band minimum to band maximum (bottom)} \label{fig:band_ratio}
\end{figure}

On the other hand, the use of a wideband dictionary does introduce the need to select the number of used bands, which directly relates to the width of the used bands as these bands are assumed to span the full parameter range. Due to the integration, the wideband elements will suffer from a reduced gain in the middle of the covered band; this will be negligible for bands of limited width, but will be pronounced, and will affect the estimation results, for wider bands. This is illustrated in Figure~\ref{fig:onebandelement}, showing a single wideband element for varying bandwidths. To quantify this effect, we define ${R}_B$ as the ratio between the minimum and maximum gains of the wideband element.
This ratio will depend on both the number of bands, $B$, and the number of samples in the signal, $N$. It may be well approximated by fitting a second order polynomial for $B$, as well as a linear fit over $N$, to the numerically computed ratios over a wide range of parameters (here, we used $B = [4,100]$ and $N = [50, 500]$), yielding
\begin{align}\label{eq:ratio}
\hat{R}_B = \frac{-0.49B^2 + 90B + 5546 - 4N}{10^4} 
\end{align}
The impact of ${R}_B$ on the resulting estimates is shown in Figure~\ref{fig:band_ratio}, where we have considered a signal  consisting of $N = 30$ samples containing $K=2$ (complex-valued) sinusoids corrupted by a zero-mean white Gaussian noise with SNR$ = 20$dB. The figure shows the percentage of correctly estimated model orders for different number of bands and the corresponding $\hat{R}_B$ ration (here, to simplify the presentation, the shown second stage zooming used a constant $B_2 = 2$ elements), computed using 1000 Monte-Carlo simulations. 
As can be seen from the figure, the percentage of correctly estimated model order will decrease as $R_b$ shrinks, with the decrease being more rapid if using fewer zooming stages. Note, however, that most of the incorrectly estimated model orders stem from over estimation and not from underestimation. Thus, it is possible to improve the order estimation in the subsequent zooming stages.
For a single stage estimator, we recommend using $\hat{R}_B > 0.81$, yielding a success rate of about 95\%; for a two stage estimator, one may reduce this further, selecting $\hat{R}_B > 0.66$ to achieve the same performance. 
After selecting an appropriate ratio for the problem at hand, taking the number of zooming steps into account, one may then determine the corresponding number of bands using \eqref{eq:ratio}, for a given $N$.
%
%

\section{Complexity analysis}\label{sect:comp}

To illustrate the computational benefits of using the wideband dictionary as compared to forming the full dictionary, we proceed with our example of determining $K$ $M$-D sinusoids by solving \eqref{eq:LASSO} using the popular ADMM algorithm \cite{BoydPCPE11_3}. In order to do so, we first transform the problem into a vector form reminiscent to \eqref{eq:vec}, and split
%
%
the variable $\boldsymbol{\beta}$ into two variables, here denoted $\x$ and $\z$, after which the optimization problem may be reformulated as
\begin{align} 
&\underset{\x, \z}{\text{min}} \,\, \frac{1}{2}||\y-\A\x||_2^2+ \lambda||\z||_1
\quad \mbox{subj. to} \quad  \x=\z \label{eq:split} 
\end{align}
%
having the (scaled) augmented Lagrangian
\begin{align}\label{eq:Lagr}
\frac{1}{2} ||\y-\A\x||_2^2+ \lambda||\z||_1+ 
\frac{\rho}{2}
 ||\x-\z+\bu||_2^2
\end{align}
where $\bu$ is the scaled dual variable and $\rho$ is the step length (see \cite{BoydPCPE11_3} for a detailed discussion on the ADMM).
The minimization is thus formed by iteratively solving \eqref{eq:Lagr} for $\x$ and $\z$, as well as updating the scaled dual variable $\bu$. 
This is done by finding the (sub-)gradient for $\x$ and $\z$ of the augmented Lagrangian, and setting it to zero, fixing the other variables to their latest values. The steps for the $j$th iteration  are thus
\begin{align}\label{eq:xstep}
\x^{(j+1)} &= \left(\A^H\A+\rho\I\right)^{-1}\left(\A^H\y+\z^{(j)}-\bu^{(j)} \right)\\
\z^{(j+1)}&= S(\x^{(j+1)}+\bu^{(j)}, \lambda/\rho)\\
\bu^{(j+1)} &= \bu^{(j)}+\x^{(j+1)}-\z^{(j+1)} \label{eq:ADMMsteps}
\end{align}
where $(\cdot)^H$ denotes the Hermitian transpose, $(\cdot)^{(j)}$ the $j$th iteration, and  $S(\bv,\kappa)$ is the soft threshold operator, defined as
\begin{align}
S(\bv , \kappa)=\frac{\text{max}\left( |\bv|-\kappa,0\right)}{\text{max}\left( |\bv|-\kappa,0\right)+\kappa}\odot \bv
\end{align}
where $\odot$ denotes the element-wise multiplication for any vector $\bv$ and scalar $\kappa$.


The computationally most demanding part of the resulting ADMM implementation is to form the inverse in \eqref{eq:xstep} and to calculate $\A^H\y$. These steps are often done by QR factorizing the inverse in \eqref{eq:xstep} prior to the iteration, so that this part is only calculated once. After this, the QR factors are used when forming the inner product. 
To give a simple example on the difference between the two types of dictionaries, we exclude any further computational speed-ups and show the difference on brute force computations of the above ADMM. This is done to give an idea on the effect  $P<N$ has on the computational complexity. The total computational cost for the step in \eqref{eq:xstep} depends on the size of the matrix $\A$.  Let $N=\prod_{m=1}^M N_m$ and $P=\prod_{m=1}^M P_m$, then $\A$ is a $N\times P$ matrix. If $P<N$, computing the inverse will cost approximately $P^3$ operations, plus an additional $P^2N$ operations to form the Gram matrix $\A^H\A$. Furthermore, to compute $\A^H\y$ requires $PN$ operations, and the final step to compute $\x$ costs $P^2$ operations. If instead $P>N$, one may make use of the Woodbury matrix identity \cite{GolubV13}, allowing the inverse to be formed using $N^3+3PN^2$ operations, whereafter one has to compute $\A^H\y$ and the final matrix-vector multiplication, together costing $PN+P^2$ operations. In total, the x-step will have the cost of roughly $P^3+(N+1)P^2+NP$, if $P<N$, or $N^3+3PN^2+PN+P^2$, if $N<P$.

Since using the banded dictionary allows for a smaller dictionary, one may calculate the computational benefit of using the integrated dictionary as compared to just using an ordinary dictionary with large $P$. Consider using only a single-stage narrowband dictionary, $\D_1$, with $P>N$ dictionary elements. This requires $C_1 = N^3+3PN^2+P^2+PN$ operations if using the above ADMM solution, with the dictionary $\D_1$ in the place of $\A$ in \eqref{eq:xstep}-\eqref{eq:ADMMsteps}. 
If, on the other hand, one uses a multiple-stage wideband dictionary with $N$ dictionary elements in the initial coarse dictionary, $\bB_1$ (which is more than required, but simplifies the calculations), the cost of forming the first stage (coarse) minimization is $C_2 = 2(N^3+N^2)$. By taking the difference, i.e., forming
\begin{align*}
R = C_1 - C_2 = N^3+3PN^2+P^2+\\+PN-2(N^3+N^2)
\end{align*}
one obtains the available computational resources, $R$, that are left for the dictionaries of the zoomed-in stages, without increasing the overall computational cost above that of the narrowband dictionary solution. Let $\bB_z$ denote the zoomed-in dictionary with $\eta N$ number of bands, where $0<\eta<1$ denotes the ratio between the number of available bands in the dictionary and the number of samples. Then, one may deduce the grid size for each $\bB_z$ that is allowed without increasing the overall computational complexity as compared to using the narrowband dictionary by solving 
\begin{align*}
R=KI_z\left((\eta N)^3+(N+1)(\eta N)^2+\eta N^2\right)
\end{align*}
where $I_z$ denotes the number of zooming steps and $K$ the number of sinusoids in the signal. 
To illustrate the resulting difference, consider the following settings: $P=1000$, $N=100$, $K=5$, and $\eta=2/3$. To only use half the resources that are needed to solve the full narrowband problem, one may, using the wideband dictionary, use $4$ stages of zooming, resulting in a grid spacing of roughly $10^{-9}$, as compared to $10^{-3}$ for the narrowband dictionary. 
One may of course also use a zooming procedure when using the narrowband dictionaries, although this would increase the risk of missing any off-grid component. This means that the smallest number of dictionary elements, for the narrowband dictionary to avoid missing any off-grid components, is $P=N$, and thus the wideband dictionary would need only at most $\eta^2$ of the computational resources needed for the ordinary dictionary, at each zooming stage. 

\begin{figure}[t]
\includegraphics[width=3.5in, height=2.5in]{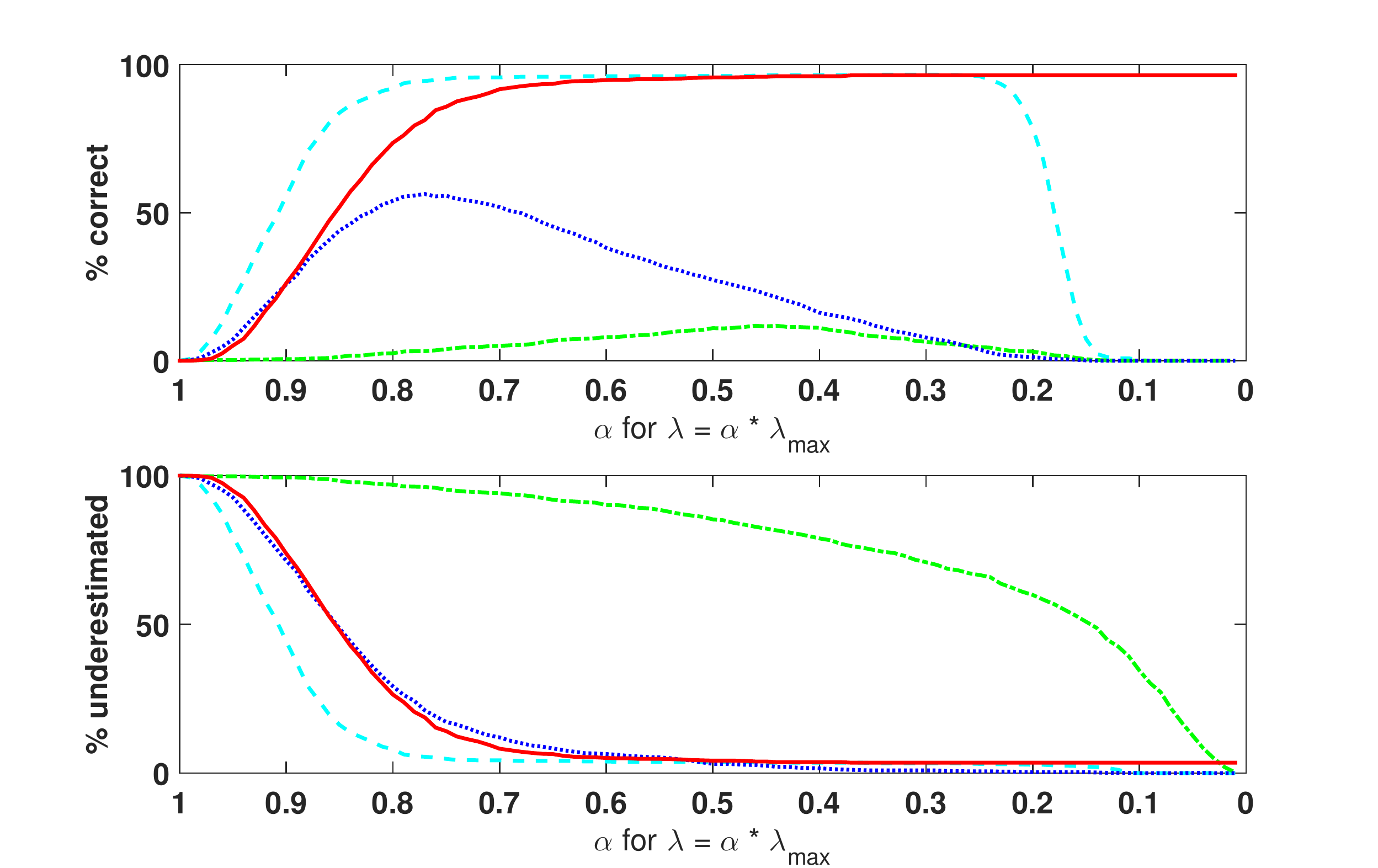}
\caption{The probability of (top) correctly estimating and (bottom) underestimating the number of spectral lines, for the (single-stage) narrowband dictionary,  using $P=1000$ elements (cyan, dashed) and $P=75$ elements (green, dot-dashed), and for the initial wideband dictionary, using $B_1=75$ elements (blue, dotted), and the  (two-stage) wideband dictionary, using $B_1=75$ elements, together with $B_2=25$ elements per activated bands in the refining dictionary (red, solid). 
%
}
\label{fig:guesses}
\end{figure}

\section{Numerical examples}\label{sect:num}

In this section, we proceed to examine the performance of the proposed method, initially illustrating that the  use of a two-stage wideband estimator will have the same estimation quality as when using the ordinary (one-stage) narrowband LASSO estimator. 

\subsection{One-dimensional data}
We initially considered a signal consisting of $N = 75$ samples containing $K=3$ (complex-valued) sinusoids corrupted by a zero-mean white Gaussian noise with SNR$ = 10$dB.
In each simulation, the sinusoidal frequencies are drawn from a uniform distribution, over $[0,1)$, with all amplitudes having unit magnitude and phases drawn from a uniform distribution over $[0,2\pi)$. The performance is then computed using three different dictionaries, namely the (ordinary) narrowband dictionary, $\D$, with $P=1000$ and $P=75$ elements, respectively, and the proposed wideband dictionary, $\bB$, using 
$B_1=75$ elements, followed by a second-stage narrowband dictionary using $B_2=25$ elements per active band. 
%
%
For each dictionary, we evaluate the performance for varying values of the user parameter $\alpha$ using 
$\lambda = \alpha\lambda_{max}$, where $\lambda_{max} = \max_i|{\bf{x}}_i^H{\bf{y}}|$ is the smallest tuning parameter value for which all coefficients in the solution are zero \cite{tibshiraniBFHSTT12_74}. Here, $\x_i$ denotes either the $i$th column of the $\D$ dictionary or the $i$th column of the $\bB$ dictionary.
Each estimated result is then compared to the ground truth, counting the number of correctly estimated and underestimated model orders. The results are shown in Figure~\ref{fig:guesses}. As can be seen from the figure, the best results are achieved when $\alpha \leq 0.65$, in which case  the proposed wideband dictionary, using $B_1=75$ bands, followed by a second stage narrowband dictionary, with $B_2=25$ for each activated band, 
have similar performance to the narrowband dictionary using $P=1000$ dictionary elements. 
 

\begin{figure}[t]
\includegraphics[width=3.5in, height=2.5in]{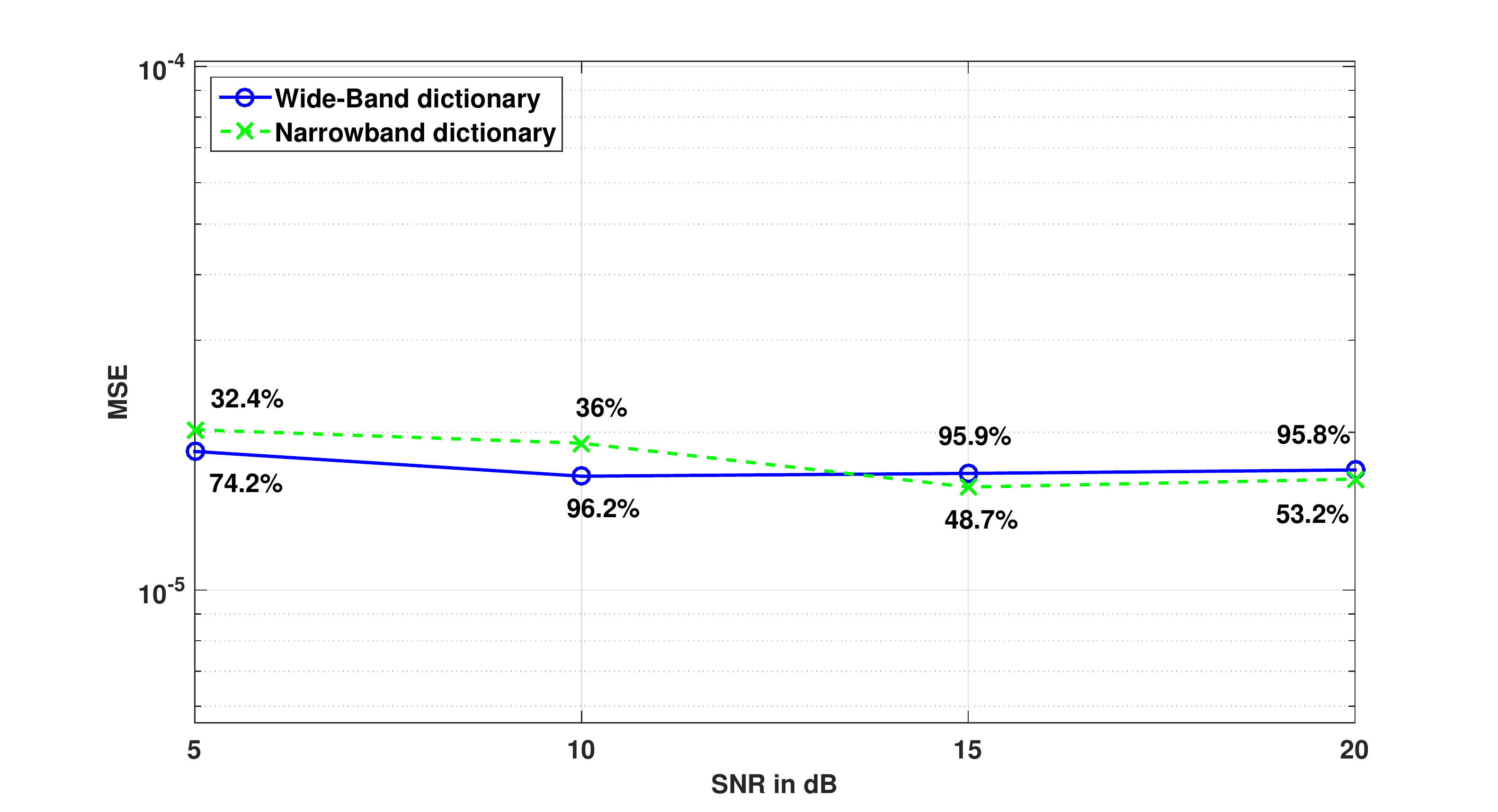}
\caption{Mean-square error curves for different SNR levels for the single-stage narrowband dictionary, using $P = 100$, as compared to the two-stage dictionary, using $B_1 = 20$ integrated wideband elements in the first stage, followed by $B_2 = 5$ wideband elements in the second stage. The percentage of correct model order estimation (excluding outliers) is shown as a percentage on top of the corresponding MSE value. 
} \label{fig:MSE_curves}
\end{figure}

\begin{figure}[t]
\includegraphics[width=3.5in, height=2.5in]{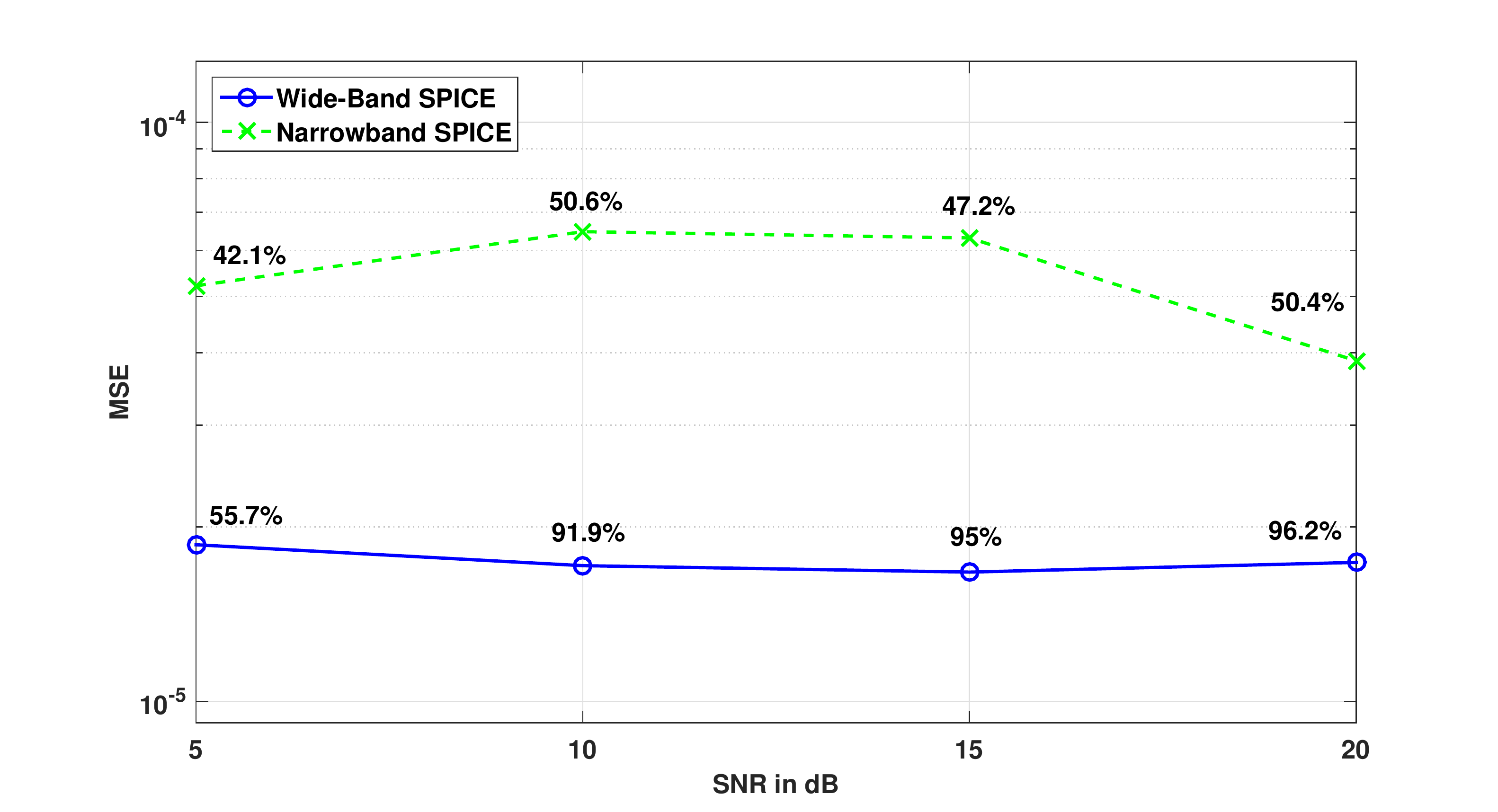}
\caption{Mean-square error curves for different SNR levels for the single-stage narrowband dictionary, using $P = 100$, as compared to the two-stage dictionary, using $B_1 = 20$ integrated wideband elements in the first stage, followed by $B_2 = 5$ wideband elements in the second stage. The percentage of correct model order estimation (excluding outliers) is shown as a percentage on top of the corresponding MSE value. 
} \label{fig:MSE_curves_spice}
\end{figure}

Proceeding, we assess the mean-square error (MSE), defined as
\begin{align} \label{MSE}
\text{MSE} = \frac{1}{K}\sum_{k = 1}^{K}(f_k - \hat{f_k})^2
\end{align}
where $f_k$ and $\hat{f_k}$ denote the true and the estimated frequency, respectively, for the two-stage dictionary, showing the MSE as a function of SNR for the first-stage wideband dictionary, $\bB_1$, and second-stage wideband refining dictionary, $\bB_2$. Here, and in the following, we consider situations where the number of elements in the dictionary is less than number of samples. As was described before, this is a situation where the performance of narrowband dictionaries can deteriorate seriously. For this experiment, we considered a signal with $N = 300$ samples containing $K = 2$ (complex-valued) sinusoids, being corrupted by different levels of zero-mean white Gaussian noise with SNR in the range $[5, 20]$ dB. 
Figure~\ref{fig:MSE_curves} shows the resulting MSE for the LASSO estimator for the estimates with correctly estimated model order; for runs with the correct model order estimation we also removed outliers from the final MSE calculation. We consider an estimate as an outlier if $|f - \hat{f}| > \Delta f$, where $\Delta f$ was defined as two times the possible resolution, where possible resolution is defined as $1/P$ for the narrowband dictionary and $1/(B_1 \cdot B_2)$ for the wideband dictionary. Figure~\ref{fig:MSE_curves_spice} shows the MSE for the same experiment done using the SPICE estimator. 
The number of outliers removed for the LASSO estimator was: $4$, $0$, $0$, $0$ for the wideband dictionary and $7$, $16$, $10$ and $11$ for the narrowband dictionary (corresponding to SNRs of $5$, $10$, $15$, and $20$ dB). The number of outliers removed for the SPICE estimator was; $17$, $1$, $1$, $0$ for the wideband dictionary and $52$, $80$, $117$, and $103$ for the narrowband dictionary. As can be seen from the figures, the two-stage dictionary using a wideband dictionary using $B_1=20$ bands, followed by a refining dictionary using $B_2=5$ wideband elements, achieves the same performance as the single-stage narrowband dictionary using $P=100$ elements in terms of resolution. However, the narrow-band dictionary will for this case fail to reliably restore the signal with reconstruction success rates of merely $30-50\%$.


Table~\ref{table:complexity} shows the relative complexity between using a full narrowband dictionary (using $P=1000$, $N=200$, and $K=2$) and some different settings for the wideband dictionaries used in the numerical section. To simplify the comparison, the given complexity is the one of solving the ADMM without utilising any structures of the dictionary matrices. 
From the table, it is clear that it is more efficient to use the zooming procedure utilising the wideband dictionary as compared to solving the same problem using a full narrowband dictionary.

\begin{table}
\begin{center}
\begin{tabular}[t]{|l||c|}
\hline
 Settings & Relative complexity\\ \hline \hline
$P=1000, N=200, K=2$   & 1 \\ \hline
$B_1=20, B_2=5$ & 0.001\\ \hline
$B_1=20, B_2=40$ & 0.015 \\ \hline
$B_1=10$, $B_2=10, B_3=5$ & 0.001\\ \hline
\end{tabular}
\caption{
Relative complexity between using the narrow- and wideband dictionaries. Here, $P$ indicates the number of columns in the narrowband dictionary, whereas $B_1$ and $B_2$ indicate the number of wideband elements in the first and second stage of the zooming procedure, respectively. In the last row, a third stage has been added using $B_3$ wideband elements.}\label{table:complexity}
\end{center}
\vspace{-5mm}
\end{table}   

\begin{figure}[t]
\includegraphics[width=3.5in, height=2.5in]{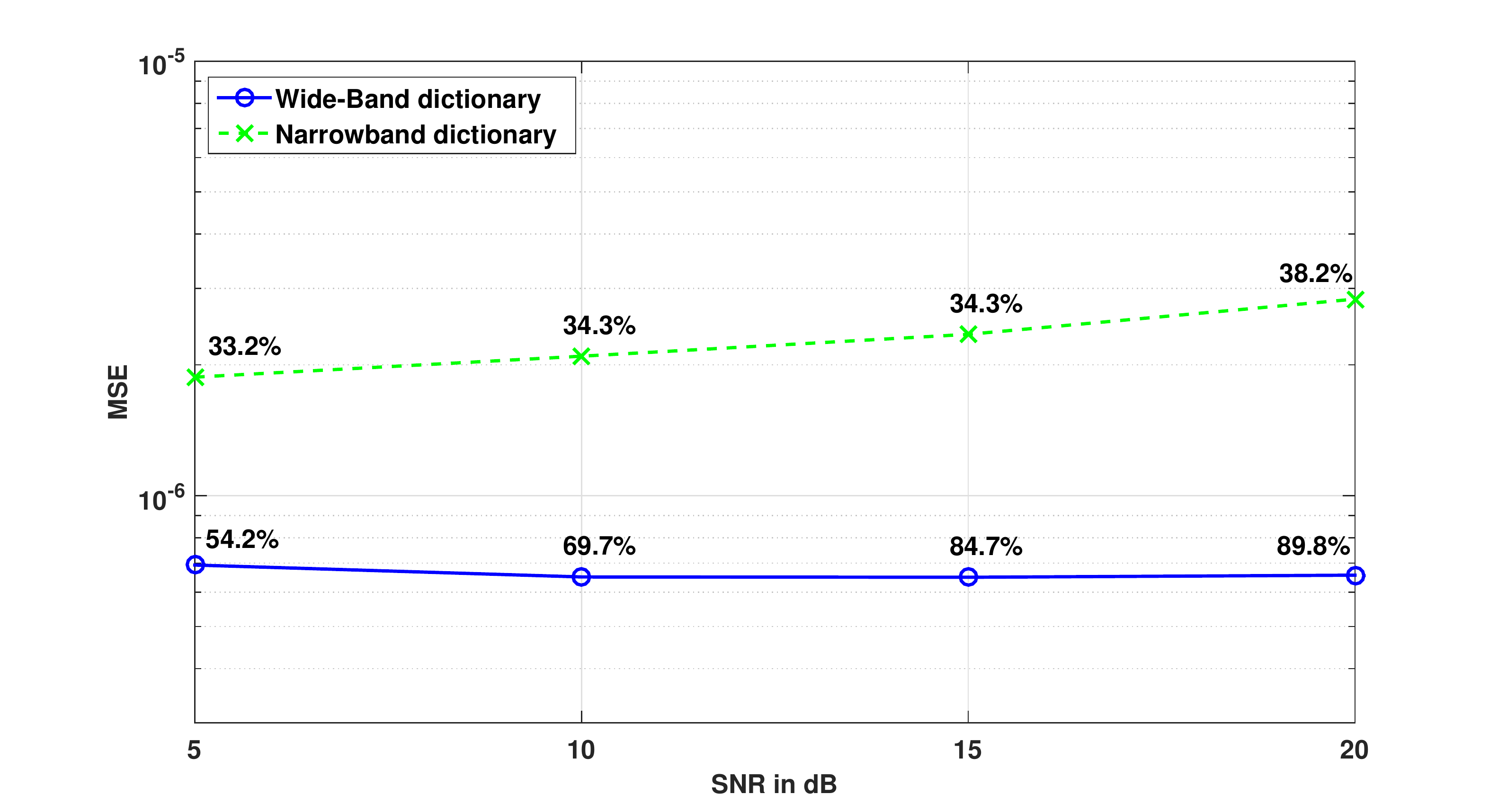}
\caption{Signal estimation for non-uniform sampling: mean-square error curves for different SNR levels for the single-stage narrowband dictionary, using $P = 200$ elements, as compared to the three-stage dictionary, using $B_1 = 10$ integrated wideband elements in the first stage, followed by $B_2 = 10$ and $B_3 = 5$ wideband dictionaries in the second stage and third stage per active band detected in the previous stage. The correct model order estimations are shown in percentage above each point.} \label{fig:1D_nonuniform}
\end{figure}

Next, we consider non-uniformly sampled data with $N = 400$ samples, for $K = 2$ sinusoids. For this experiment, we also added a third estimation step for the iterative wideband dictionary. After initial estimation with $B_1 = 10$ wideband dictionary elements, we zoom into the active bands with $B_2 = 10$ dictionary elements per active band, and then once again with $B_3 = 5$ dictionary elements. In spite of the three stage zooming, the method requires considerably less computational operations as compared to using a corresponding narrowband dictionary, but results in better performance both in terms of resolution and model-order accuracy. The resulting MSEs are shown in Figure~\ref{fig:1D_nonuniform}. All results are computed using 1000 Monte-Carlo simulations. 

\begin{figure}[t]
\includegraphics[width=3.5in, height=2.5in]{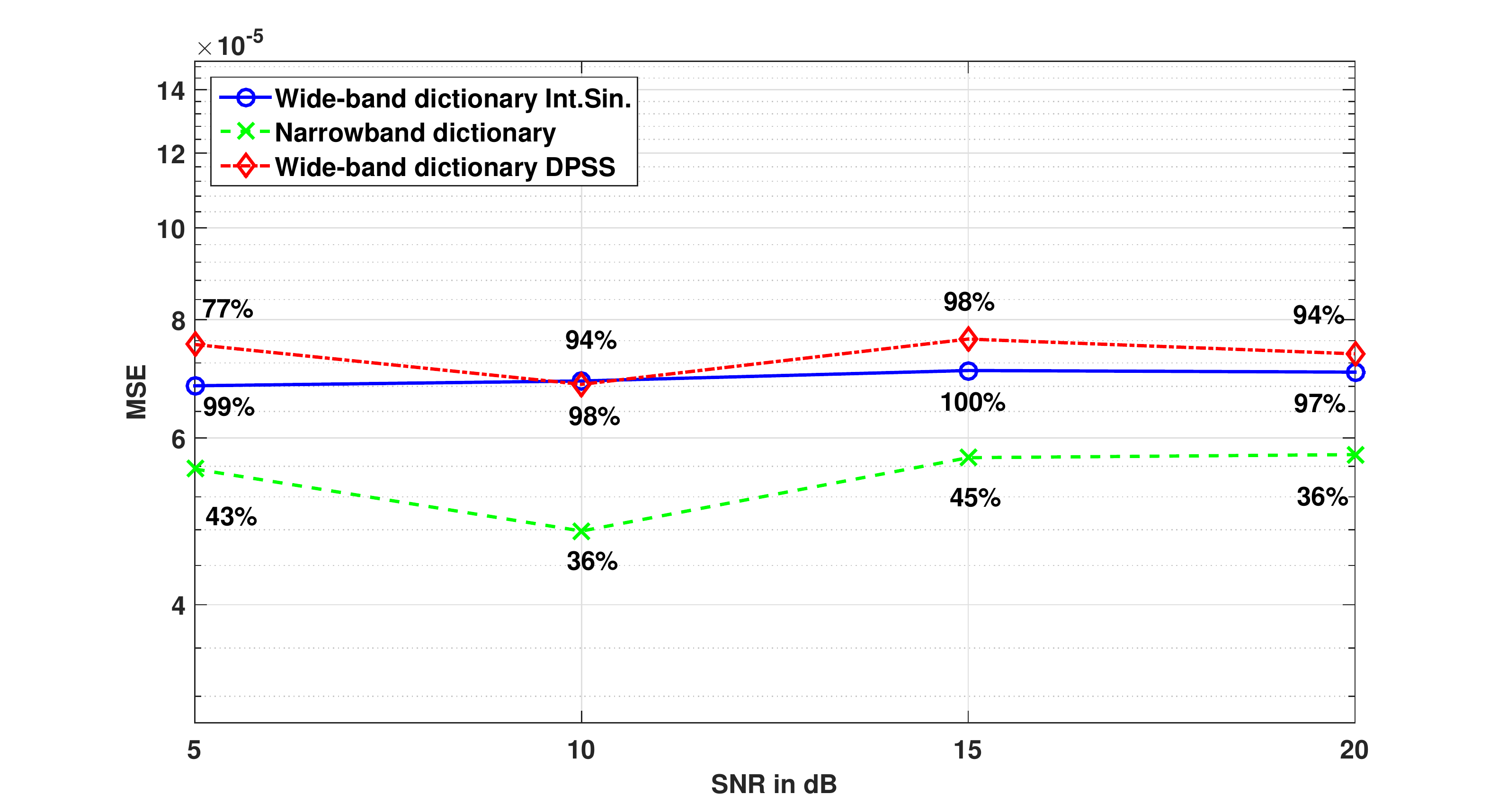}
\caption{Signal estimation in two dimensions: mean-square error curves for different SNR levels for the single-stage narrowband dictionary $\D$, using $P = 49$ per dimension, as compared to the two-stage dictionaries (DPSS based and integrated sinusoids based), using $B_1 = 7$ wideband elements in the first stage, followed by $B_2 = 7$ wideband elements in the second stage (per active band).} \label{fig:MSE_2D}
\end{figure}

\subsection{Two-dimensional data}
In this subsection, we present results on a $2$-D data set. In this example, each dimension is sampled uniformly with $N = 100$ samples. We compare a narrowband dictionary with $P = 49$ elements per dimension with the wideband dictionary using $B_1 = 7$ bands per dimension in the first step and a wideband dictionary with $B_2 = 7$ elements per active band in a second (zooming) step. Here, we use two separate wideband dictionaries, the first, $\bB$, using integrated dictionary elements as defined in \eqref{eq:prop}, and the second, $\bB_{DPSS}$, which contains elements based on DPSS. For the DPSS-based dictionary, we used a sequence length of $Q=100$ and $W=1/2.1$. Using $W<1/2.1$ results in dictionary elements which concentrate energy in a more narrow band and are therefore not suitable for the dictionary with $B_1 = B_2 = 7$ elements. We considered a signal containing $K=2$ (complex-valued) sinusoids per dimension, with the signal being corrupted by a zero-mean white Gaussian noise. 
In each simulation, the sinusoidal frequencies are drawn from a uniform distribution, over $[0,1)$, with all the amplitudes having unit magnitude. The two dictionaries are compared against each other based on the MSE performance in the same manner as in the previous subsection, with the MSE being calculated as the average value for both dimensions if the model order estimate for the iteration was correct. 
The percentages of correct model order estimates are shown for each SNR value.
Figure~\ref{fig:MSE_2D} shows the resulting MSE curves. It can be seen that the wideband dictionary with integrated sinusoids marginally outperforms the DPSS-based wideband dictionary both in terms of MSE and model-order accuracy. Comparing to using the narrowband dictionary, it can be seen that both wideband dictionaries outperform it considerably in terms of model-order estimation, although the narrowband dictionary shows slightly better performance in terms of MSE. Also in this example, the wideband dictionaries provide a considerable reduction in computational complexity as well as a robustness in terms of estimating off-grid components. All results are computed using 100 Monte-Carlo simulations.

\begin{figure}[t]
\includegraphics[width=3.5in, height=2.5in]{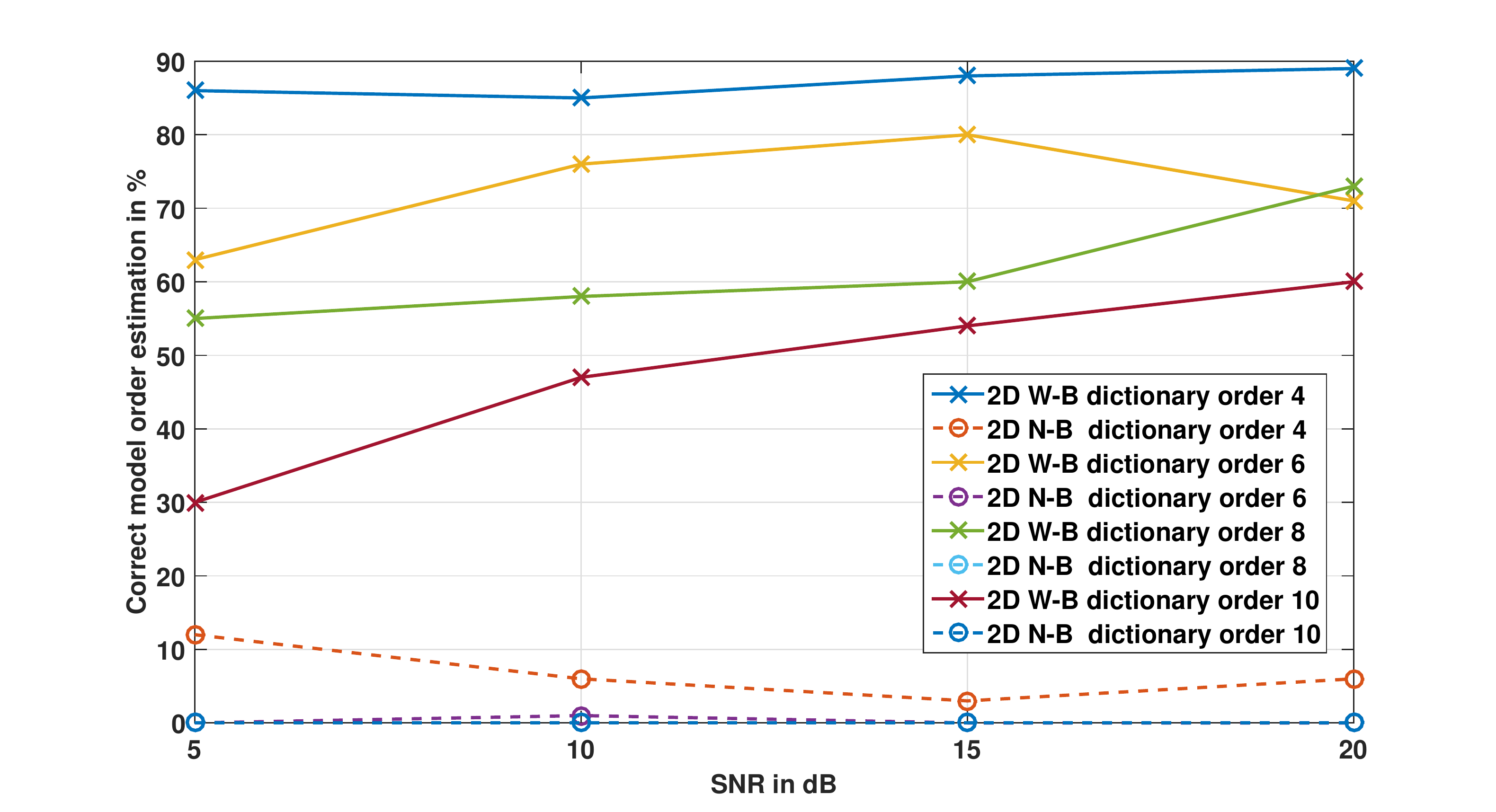}
\caption{Percentage of correct model order esimations for different number of sinusoids and for different SNR levels for wideband dictionary (W-B) and narrowband dictionary (N-B).} \label{fig:model_order}
\end{figure}

Using the same setup as described above we also evaluated the performance of the proposed approach when the number of sinusoids to detect is higher. Again, we considered the ordinary narrowband dictionary, $\D$, and the wideband dictionary, $\bB$, from the previous experiment. We calculated the percentage of correct model order estimation for signals with $K = 4, 6, 8$, and $10$ (complex-valued) sinusoids. The results were computed using $100$ Monte-Carlo simulations; the correct model order estimation percentages for different SNR levels are shown in Figure~\ref{fig:model_order}. The best regularization parameters $\lambda$ for solving the LASSO for each case were found beforehand with the grid-search method. For this, we selected the range of parameter $\alpha \in [0.7,0.05]$ with the step-size 0.05 and ran 100 Monte-Carlo simulations for each model order and then picked the best parameter for the selected model order based on model order accuracy. For the two-step wideband dictionary, a grid-search was done for the set of $\alpha$ parameter for the both stages. It can be clearly seen that for situations where the number of elements in the dictionary is lower than the number of samples, the narrow-band dictionary fails to produce any meaningful results.

\begin{figure}[t]
\includegraphics[width=3.5in, height=2.5in]{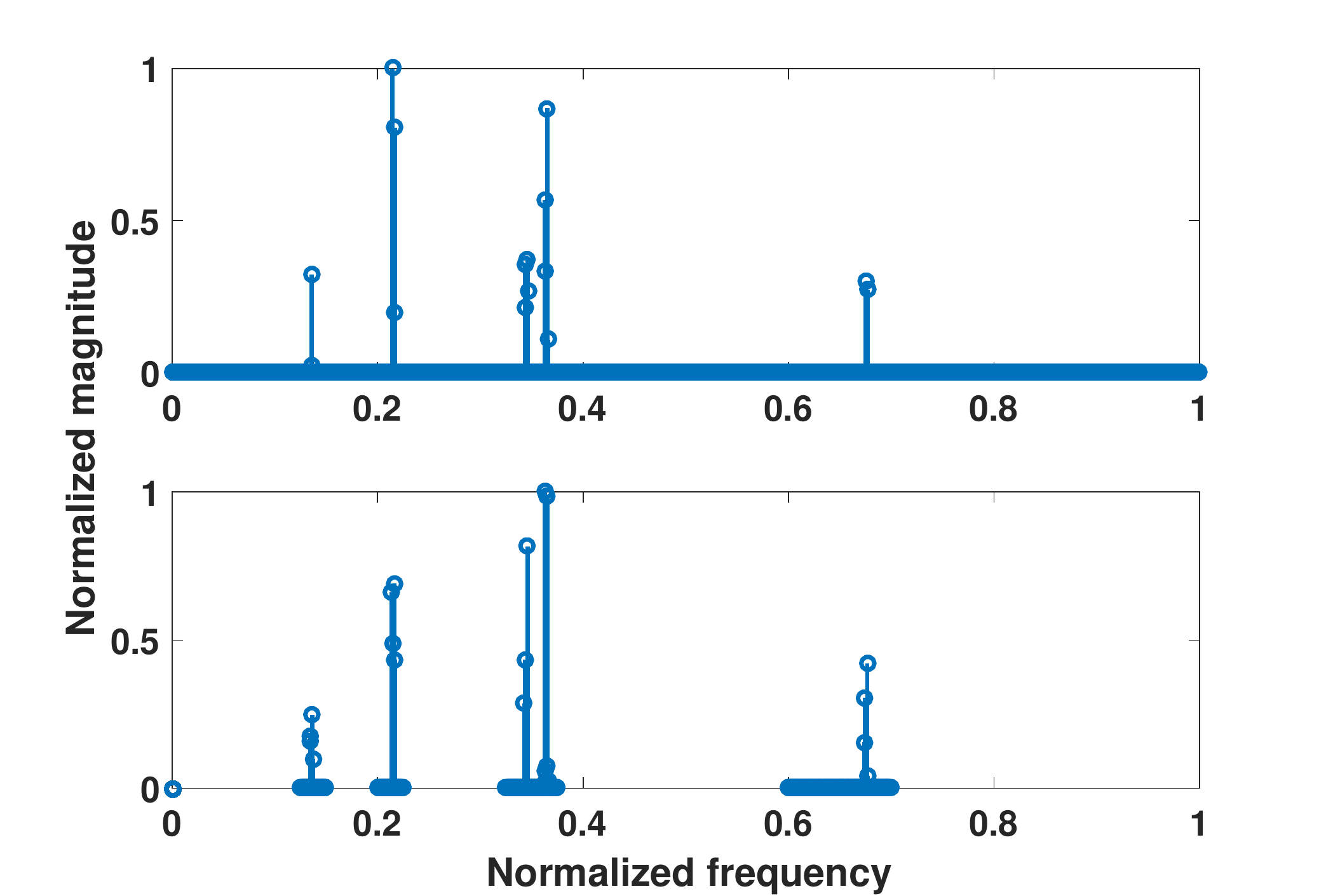}
\caption{The resulting estimates using a dictionary with $2000$ narrowband elements (top), and a two-stage zooming approach using wideband elements, using $B_1=40$ in the first stage and $B_2=50$ for each activated bands in a second stage. The signal is a measured NMR signal of length $N=256$. }\label{fig:realNMRDataExample}
\end{figure}

\subsection{Measured data example}
Finally, we examine the performance of the proposed wideband framework on measured nuclear magnetic resonance (NMR) data, again comparing with using the full narrowband dictionary. The measured data 
NMR measurement consist of $N=256$ samples, and contains five damped sinusoidal signals. To make the comparison fair, we neglect the damping in the modelling (as the wideband dictionary will implicitly allow for the resulting wide peaks, whereas the narrowband dictionary will require an additional parameter to do so). This results in estimates containing clusters of peaks instead of individual component. 
Figure~\ref{fig:realNMRDataExample} illustrates the resulting estimates, showing the result of using a narrowband dictionary with 2000 elements (top), as well as a two-stage wideband dictionary (using $B_1=40$ elements in the first stage and $B_2=50$ elements in the second). The resulting estimates will thus have the same final grid resolution. 
As can be expected, both estimators show similar results, having the same support and roughly the same relative amplitudes. Using the introduced ADMM implementation described in Section~\ref{sect:comp}, the wideband estimate was formed in $0.315$ seconds, which was $20$ times faster than the narrowband estimate. Here, one may note that if an iterative narrowband zooming would be used, it would require at least $256$ elements in the first stage to avoid losing any peaks; doing so would require more complexity than the two-stage wideband estimator.

\section{Conclusion}

In this paper, we have introduced a wideband dictionary framework, allowing for a computationally efficient reconstruction of sparse signals. Wideband dictionary elements are formed as spanning bands of the considered parameter space. In the first stage, one may typically use a coarse grid using the integrated wideband dictionary locating the bands of interest, whereafter non-active parts of the parameter space are discarded. In the next stage, a refining dictionary can be used to more precisely determine the parameters of interest on the active bands from the previous step, allowing for an iterative zooming procedure. The technique is illustrated for the problem of estimating multidimensional sinusoids corrupted by Gaussian noise, showing that the same accuracy can be achieved, although at a computationally substantially lower cost and with much less risk of missing any off-grid components. The proposed framework is here illustrated for the LASSO and SPICE estimators, but other sparse reconstruction techniques may be extended similarly. 
\vspace{-.1 cm}

\bibliographystyle{IEEEbib}
\bibliography{IEEEabrv,referencesAll}

\end{document}